\documentclass[prb,reprint,amsmath,amssymb,superscriptaddress]{revtex4-2}
\usepackage{graphicx}
\usepackage{dcolumn}
\usepackage{bm}
\usepackage{textcomp}
\usepackage[colorlinks= true,urlcolor=blue,linkcolor= blue,citecolor=blue,bookmarks=false,pdfstartview=]{hyperref}

\begin{document}
\title{Field-orientation-dependent magnetic phases in GdRu$_2$Si$_2$ probed with muon-spin spectroscopy}
\author{B.\,M. Huddart}
\affiliation{Department of Physics, Durham University, Durham DH1 3LE, United Kingdom}
\affiliation{Clarendon Laboratory, University of Oxford, Department of Physics, Oxford OX1 3PU, United Kingdom}
\author{A. Hern\'{a}ndez-Meli\'{a}n}
\affiliation{Department of Physics, Durham University, Durham DH1 3LE, United Kingdom}
\author{G.\,D.\,A. Wood}
\affiliation{University of Warwick, Department of Physics, Coventry CV4 7AL, United Kingdom}
\author{D.\,A. Mayoh}
\affiliation{University of Warwick, Department of Physics, Coventry CV4 7AL, United Kingdom}
\author{M. Gomil\v{s}ek}
\affiliation{Jo\v{z}ef Stefan Institute, Jamova c. 39, SI-1000 Ljubljana, Slovenia}
\affiliation{Faculty of Mathematics and Physics, University of Ljubljana, Jadranska u. 19, SI-1000 Ljubljana, Slovenia}
\author{Z. Guguchia}
\affiliation{PSI Center for Neutron and Muon Sciences CNM, 5232 Villigen PSI, Switzerland}
\author{C. Wang}
\affiliation{PSI Center for Neutron and Muon Sciences CNM, 5232 Villigen PSI, Switzerland}
\author{T. J. Hicken}
\affiliation{PSI Center for Neutron and Muon Sciences CNM, 5232 Villigen PSI, Switzerland}
\author{S.\,J.\,Blundell}
\affiliation{Clarendon Laboratory, University of Oxford, Department of Physics, Oxford OX1 3PU, United Kingdom}
\author{G. Balakrishnan}
\affiliation{University of Warwick, Department of Physics, Coventry CV4 7AL, United Kingdom}
\author{T. Lancaster}
\affiliation{Department of Physics, Durham University, Durham DH1 3LE, United Kingdom}
\date{\today}

\begin{abstract}
Centrosymmetric GdRu$_2$Si$_2$  exhibits a  variety of multi-Q magnetic states as a function of temperature and applied magnetic field, including a square skyrmion-lattice phase.  The material's behavior is strongly dependent on the direction of the applied field, with different phase diagrams resulting for fields applied parallel or perpendicular to the crystallographic $c$ axis. Here, we present the results of muon-spin relaxation ($\mu^+$SR) measurements on single crystals of GdRu$_2$Si$_2$. 
Our analysis is based on the computation of muon stopping sites and consideration of zero-point motion effects, allowing direct comparison with the underlying spin textures in the material. The muon site is confirmed experimentally, using angle-dependent measurements of the muon Knight shift. Using transverse-field $\mu^+$SR with fields applied along either the [001] or [100] crystallographic directions, we distinguish between the magnetic phases in this system via their distinct muon response, providing additional evidence for the skyrmion and meron-lattice phases, while also suggesting the existence of RKKY-driven muon hyperfine coupling. Zero-field $\mu^+$SR provides clear evidence for a transition between two distinct magnetically-ordered phases at 39~K.
\end{abstract}

\maketitle

\section{Introduction}

Skyrmions are topologically-protected vortices of magnetization that  exist in some magnetically-ordered materials \cite{doi:10.1080/00107514.2019.1699352}.
 The search for new materials that host a magnetic skyrmion state has been motivated by the numerous potential applications of skyrmions in the areas of ultra-efficient information storage and spintronics \cite{Fert2017}. Skyrmions are typically stabilized by the Dzyaloshinskii--Moriya interaction (DMI), which can only occur in systems
that lack inversion symmetry. Magnets based on Gd (possessing 4f$^7$ core states with $S=7/2$, $L=0$) are promising candidates for realizing non-collinear spin structures such as skyrmions, since the moments show non-preferential orientation. Two Gd-based magnets where skyrmions have recently been discovered
are Gd$_2$PdSi$_3$ \cite{doi:10.1126/science.aau0968} and Gd$_3$Ru$_4$Al$_{12}$ \cite{Hirschberger2019}. In these centrosymmetric materials, skyrmions have been proposed to be stabilized by either short- or long-range magnetic frustration rather than DMI. This can lead to interesting competition with other magnetic phases, and potentially
new behavior of the skyrmion state.
Skyrmions have also been observed in GdRu$_2$Si$_2$~\cite{Khanh2020}, another centrosymmetric Gd-based magnet. However, the structure of GdRu$_2$Si$_2$, shown in Fig.~\ref{fig:sites}(a), comprises alternating square-lattice Gd and Ru$_2$Si$_2$ layers and therefore, unlike for Gd$_2$PdSi$_3$ and Gd$_3$Ru$_4$Al$_{12}$, short-range geometrical frustration cannot be used to explain the formation of the skyrmion lattice (SkL). Magnetic resonant x-ray scattering (RXS) and Lorentz transmission electron microscopy reveal the existence of a double-Q magnetic state, corresponding to a square SkL \cite{Khanh2020}. This is distinct from the hexagonal SkL found in Gd$_2$PdSi$_3$ and GdRu$_4$Al$_{12}$, and in the vast majority of the noncentrosymmetric skyrmion-hosting materials. Moreover, the existence of a square SkL has only been demonstrated \cite{Khanh2020} for applied fields $\boldsymbol{B}_0 \parallel$ [001],   
for which the SkL phase is stabilized at $B_0\approx$ 2 T. For fields in the $ab$ plane, a different phase diagram is obtained \cite{GARNIER199680,https://doi.org/10.1002/advs.202105452}. The magnetic phase diagrams for $\boldsymbol{B}_0 \parallel$ [001] and $\boldsymbol{B}_0 \parallel$ [100] have been studied using RXS \cite{https://doi.org/10.1002/advs.202105452}, with these measurements revealing a rich variety of double-Q magnetic structures, including the antiferroic order of meron/antimeron-like textures. Neutron diffraction measurements on single-crystal and polycrystalline $^{160}$GdRu$_2$Si$_2$ found, in addition to scattering at the principal propagation vectors $\boldsymbol{q}_1$ and $\boldsymbol{q}_2$, higher-order satellites at $\boldsymbol{q}_1+2\boldsymbol{q}_2$. This prompted a revision of the zero-field magnetic ground state to a double-Q constant-moment magnetic structure, with one-dimensional topological charge density \cite{PhysRevB.107.L180402}. At small values of applied field, an additional phase has been identified just below $T_{\mathrm{N}}$ whose properties are less well understood, but which was recently revealed to involve a single magnetic propagation vector that splits into two wave vectors with different magnitudes upon further cooling~\cite{paddison2024spindynamicscentrosymmetricskyrmion}.

\footnotetext{See Supplemental Material at [URL will be inserted by publisher] for further details of the the density functional theory calculations and simulations of the magnetic field distributions, which includes Refs.~\cite{PhysRevB.20.850,wimda,minuit,iminuit,CASTEP,PBE,HIEBL1983287,mpgrid1976,mufinder,magres_review,govind, NEB,kantorovich,griffiths,Allodi_2014}.}

Here, we use muon-spin spectroscopy ($\mu^+$SR)~\cite{muon_book} to probe the magnetic phases in GdRu$_2$Si$_2$ as a function of magnetic field and temperature. The $\mu^+$SR technique has been shown to be sensitive to the distribution of magnetic fields arising from the SkL phase in Cu$_2$OSeO$_3$ \cite{PhysRevB.91.224408}, as well as those found in the helical \cite{PhysRevB.93.144419} and conical \cite{PhysRevB.95.180403} phases of skyrmion-hosting MnSi. Implanted muons have also been shown to effectively probe the emergent dynamics associated with Bloch SkL, such as in Cu$_2$OSeO$_3$ and Co$_x$Zn$_y$Mn$_{20-x-y}$ \cite{PhysRevB.103.024428} and in Gd$_2$PdSi$_3$~\cite{gomilsek2023anisotropic}, or those accompanying a N\'{e}el SkL, as found in GaV$_4$S$_8$ and GaV$_4$Se$_8$ \cite{PhysRevB.98.054428,PhysRevResearch.2.032001}.
Using transverse-field (TF) $\mu^+$SR for magnetic fields applied along either [001] or [100], we show here how the response of the implanted muon reflects the occurrence of skyrmion and other double-Q magnetic phases.  
Zero-field (ZF) $\mu^+$SR measurements provide clear evidence for a transition between two distinct magnetically-ordered phases at 39~K, showing that the phase reported just below $T_{\mathrm{N}}=45$~K is ordered throughout the bulk of the material. Our modelling, which includes muon-site determination and its confirmation via experiment, also provides an estimate of the low-temperature ordered moment of $4.8(1)\mu_{\mathrm{B}}$ and shows the influence of long-range Ruderman-Kittel-Kasuya-Yosida (RKKY) coupling in the system, via the muon's coupling to the electronic degrees of freedom.

The paper is structured as follows: in Sec.~\ref{sec:muon_sites} we present the results of muon stopping site calculations using density functional theory (DFT), in Sec.~\ref{sec:results} we present the results of $\mu^+$SR measurements on GdRu$_2$Si$_2$ and discuss these with support from simulations of the magnetic field at the muon site for candidate magnetic structures, and finally we present our conclusions in Sec.~\ref{sec:conclusion}. Additional details can be found in the Supplemental Material (SM)~\cite{Note1}.

\section{Muon site calculations}\label{sec:muon_sites}

\begin{figure}[ht]
	\centering
	\includegraphics[width=\columnwidth]{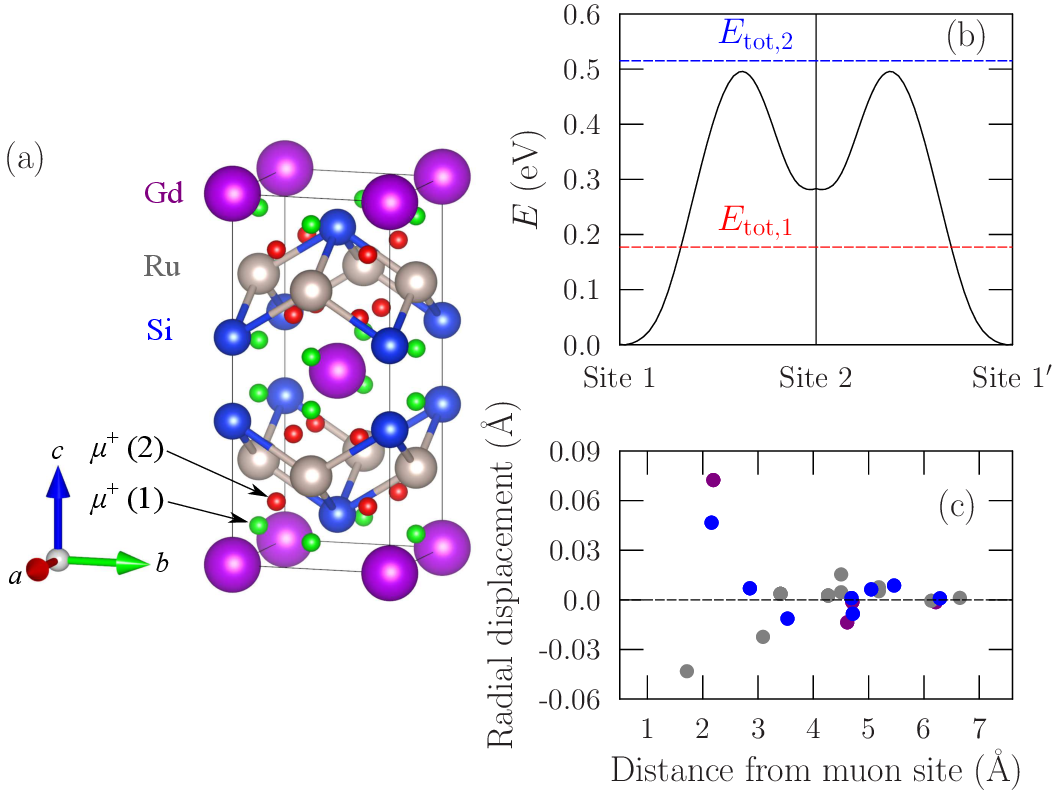}
	\caption{(a) Two crystallographically distinct muon stopping sites in GdRu$_2$Si$_2$, obtained using DFT. (b) Energy barriers between muon  sites. Red and blue lines indicate the total (classical plus quantum, zero-point) energy of the muon at sites 1 and 2, respectively. (c) Muon-induced displacements of ions as a function of distance from the lowest-energy muon site.}
	\label{fig:sites}
\end{figure}

Implanted muons are sensitive to the distribution of magnetic fields at the muon stopping sites. Knowledge of these sites can facilitate a  quantitative understanding of $\mu^+$SR measurements. Progress has recently been made in computing stopping sites using DFT \cite{DFTplusMu}, although treatment of quantum-mechanical effects such as zero-point motion (ZPM) has hitherto been limited to special cases~\cite{gomilsek2023many,muon_book}. We have used DFT methods to calculate the muon sites in GdRu$_2$Si$_2$ \cite{Note1} and find two crystallographically-distinct candidate  sites [Fig.~\ref{fig:sites}(a)]. These sites were further investigated to evaluate their stability. In the weakly-bound adiabatic limit (i.e., with zero entanglement with nearby nuclear positions)~\cite{gomilsek2023many}, the zero-point energy (ZPE) of the muon can be approximated as the sum of the three highest frequency phonon modes. The entanglement witness $w_1$, obtained by projecting the three highest-energy (normalized) phonon normal modes onto pure muon motion, summing the squared norms of the resulting (projected) phonon eigenvectors, and subtracting the value 3~\cite{gomilsek2023many}, can be used to determine whether the muon is in fact in this limit. This results in $w=0$ for zero muon–nuclear entanglement (weakly-bound adiabatic limit) and $w_1 < 0$ in the entangled, many-body case. We obtained entanglement witnesses $w_1=-0.0079$ and $w_1=-0.0082$ for site 1 and site 2, respectively, which are both close to zero, confirming the validity of the weakly-bound adiabatic approximation, i.e., the decoupling of muon vibrational modes from the vibrational modes of the lattice. 

The energy barriers for moving a muon between sites 1 and 2 were calculated using transition state searches, and confirmed using the nudged elastic band (NEB) method \cite{NEB}, yielding the energy profile in Fig.~\ref{fig:sites}(b). However, on taking into account the quantum zero-point energy, a muon at site 2 (the higher-energy state of the two) has sufficient energy to overcome the energy barrier between its classical location and that of site 1. Site 2 is therefore not stable under quantum fluctuations of the muon position. 

While the barrier from site 1 to site 2, and hence the barrier between adjacent instances of site 1 (i.e. site 1 and site $1'$), is sufficiently high that we do not expect muons at site 1 to be able to leave this site classically, we also need to consider the possibility of quantum mechanical tunneling. By considering the effective one-dimensional potential for the muon along the minimum-energy path between the classical turning points, i.e., where the black and red lines in Fig.~\ref{fig:sites}(b) intersect, we used the WKB approximation to estimate a tunneling rate $\approx$0.05~Hz for muons to move between site 1 and site $1'$~\cite{Note1}, which is significantly longer than the timescale over which measurements are made. We therefore conclude that site 1 is the single crystallographically-distinct stable stopping site in this material and we plot the displacements of the nearby ions in the vicinity of this site in Fig.~\ref{fig:sites}(c). The muon does not introduce any significant perturbations to its local environment, with all displacements $<0.1$~\AA{}, and is therefore expected to be a faithful probe of the magnetism in this system. Note that these calculations do not take into account the possible effect of charge density waves, which have been proposed to result from the coupling between itinerant electrons and localized moments in this system~\cite{yasui}, and could result in small perturbations to the muon site.

\section{Results and Discussion}\label{sec:results}

\subsection{Angle-dependent Knight shift measurements}\label{sec:knight_shift}
To further examine the muon site and its associated hyperfine coupling, we carried out angle-dependent measurements of the muon Knight shift on a single crystal of GdRu$_2$Si$_2$ with a large [100] face using the GPS spectrometer at the Swiss Muon Source, Paul Scherrer Institute (Switzerland). Measurements were made at 55~K, i.e., within the paramagnetic phase. The sample was enclosed in aluminum foil and attached to the end of a sample stick. 
These measurements were made in a spin-rotated configuration, with the initial muon-spin polarization making an angle $\approx 45^\circ$ to the muon beam axis.

\begin{figure}[htb]
 	\centering
 	\includegraphics[width=\columnwidth]{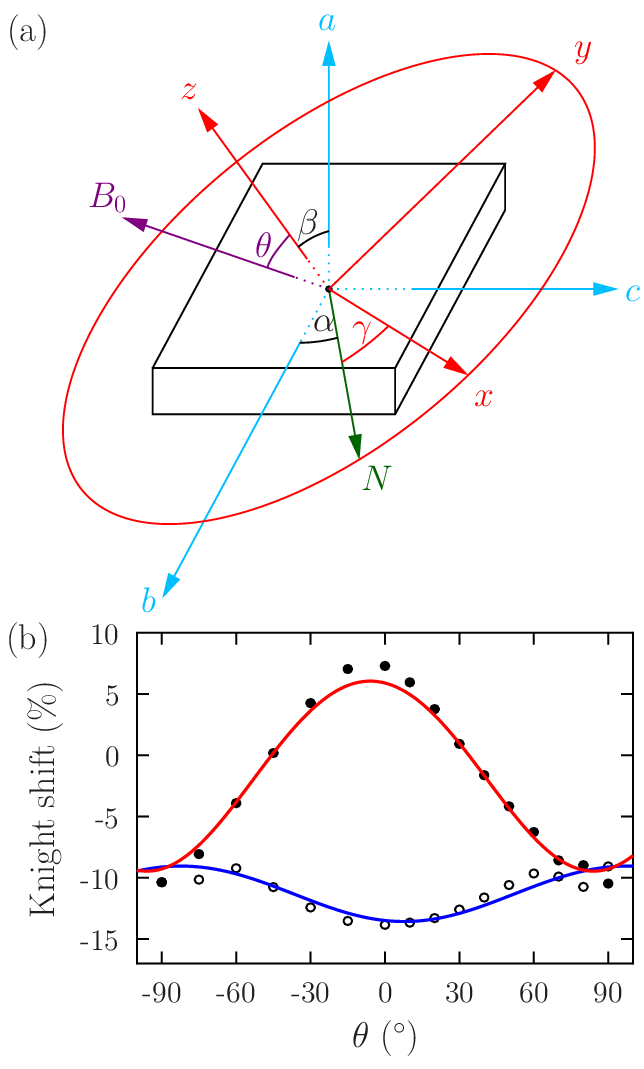}
 	\caption{(a) Schematic of the experimental geometry, indicating the direction of the applied magnetic field $\boldsymbol{B}_0$ in the frame ($a$,$b$,$c$) of the sample. (b) Two components of the muon Knight shift as a function of the rotation angle $\theta$.}
 	\label{fig:ks}
 \end{figure}

The experimental geometry is shown in Fig~\ref{fig:ks}(a). For the purpose of calculating the magnitude of magnetic field at the muon site, we consider the crystallographic frame ($a$, $b$, $c$) of the sample to be fixed. We then define a second frame, the lab frame, such that the applied magnetic field $\boldsymbol{B}_0$, with magnitude 0.78~T, initially points along $\boldsymbol{\hat{z}}$ and rotates anticlockwise about $\boldsymbol{\hat{x}}$, with the angle of rotation denoted $\theta$. The two coordinate systems are related using the Euler angles $\alpha$, $\beta$, and $\gamma$, as shown in Fig~\ref{fig:ks}(a). Assuming perfect alignment of the crystal, i.e., $\alpha=\beta=\gamma=0$, we have $\boldsymbol{\hat{z}}=\boldsymbol{\hat{a}}$ and $\boldsymbol{\hat{x}}=\boldsymbol{\hat{b}}$. This corresponds to the scenario in which the applied magnetic field is initially along [100] and the rotation axis of the sample stick is along [010]. The use of Euler angles allows us to capture any sample misalignment in our analysis.

At each rotation angle, we consider the components of the muon polarization measured in the four detectors perpendicular to the applied magnetic field: up, down, left, and right. These are all fit to a function
\begin{equation}\label{eq:tf_time_domain}
A(t) = A_\mathrm{bg} \cos(\omega_0 t + \phi) + \sum_{i=1}^2 A_i e^{-\lambda_i t} \cos(\omega_i t + \phi),
\end{equation}
where the first term is due to muons stopping outside the sample and precessing about the applied field, while the other two components correspond to muons stopping in the sample. The phase $\phi$ is allowed to vary for each detector, but is the same for all components in the fitting function, as would be expected for the magnetic field distribution arising from a field-polarized state. Note that it was not possible to separately resolve $\omega_1$ and $\omega_2$ for $\theta = -90^\circ$ and hence these were fixed to be equal in the fitting procedure for this angle (as were $\lambda_1$ and $\lambda_2$).

For muons at stopping at site 1 we would expect to observe two distinct components with frequencies shifted away from the Larmor frequency $\omega_0=\gamma_\mu B_0$ corresponding to the applied field. For Gd moments pointing in a general direction within the $ac$ plane, instances of site 1 lying on the $a$ axis will experience a different dipolar magnetic field to those lying on the $b$ axis. As each of the magnetically distinct instances of site 1 are expected to be occupied with equal probability, we fix $A_1=A_2$ in Eq.~\eqref{eq:tf_time_domain}. Furthermore, from a fit to the spectrum measured for $\theta = 0$, where the frequency shifts are at their largest, we find $(A_1+A_2)/(A_1+A_2+A_\mathrm{bg})=0.95(2)$. The ratio of the amplitudes is fixed at this value for fits at the other rotation angles. For each component, the precession frequency is related to the magnetic field at the muon site through $\omega_i =\gamma_\mu B_i$. Hence, we can compute the Knight shift at each of the muon sites through
\begin{equation}
K_i = \frac{\omega_i - \omega_0}{\omega_0}.
\end{equation}

The muon Knight shifts corresponding to the two components of the asymmetry are shown as a function of direction of the applied magnetic field in Fig.~\ref{fig:ks}(b).  These two components exhibit the maximum separation for $\theta \approx 0$ and coincide when $\theta$ is close to 90$^\circ$. This can be understood in the case of perfect alignment of the crystal, for which a value of $\theta=90^\circ$ would correspond to the applied field, and hence the Gd$^{3+}$ moments, pointing along [00$\bar{1}$]. In that case, all instances of site 1 would become magnetically equivalent, and hence only a single Knight shift would be observed.

In this experiment, the contributions to the magnetic field at the muon site are as follows
\begin{equation}
    \boldsymbol{B}(\theta) = \boldsymbol{B}_0 + \boldsymbol{B}_\mathrm{dip}(\theta) + \boldsymbol{B}_\mathrm{Lor} + \boldsymbol{B}_\mathrm{cont} + \boldsymbol{B}_\mathrm{demag}(\theta),
\end{equation}
where $\boldsymbol{B}_0$ is the applied magnetic field, $\boldsymbol{B}_\mathrm{dip}(\theta)$ is the local dipolar field at the muon site, $\boldsymbol{B}_\mathrm{Lor}$ is the Lorentz field, $\boldsymbol{B}_\mathrm{cont}$ is the contact-hyperfine field and $\boldsymbol{B}_\mathrm{demag}(\theta)$ is the demagnetizing field.
The Lorentz term $\boldsymbol{B}_\mathrm{Lor}$ is proportional to the sample magnetization.
We assume that the magnetic susceptibility is independent of field direction within the paramagnetic phase. Hence, we assume Gd$^{3+}$ moments with a magnitude $\mu_\mathrm{Gd}$ in the direction of the applied field, where $\mu_\mathrm{Gd}$ is allowed to vary in the fitting, but is assumed to have the same value for all orientations of the applied field. Using this magnetic structure, we can calculate $\boldsymbol{B}_\mathrm{Lor}$, and, using the dipolar tensor at the muon stopping site, the local dipolar field $\boldsymbol{B}_\mathrm{dip}$ at each field angle. 

The contact term $\boldsymbol{B}_\mathrm{cont}$ was evaluated using 
the approach implemented in MuESR~\cite{doi:10.7566/JPSCP.21.011052}, whereby the hyperfine field is assumed to be isotropic and results from a scalar coupling to the magnetic moments. Each of the nearest-neighbor moments contributes to the total hyperfine field by an amount inversely proportional to the cube of its distance from the muon. This field is then scaled by the hyperfine coupling constant $A$, leading to
\begin{equation}\label{eq:hyperfine}
    \boldsymbol{B}_\mathrm{cont} = A \frac{2\mu_0}{3} \sum_{i=1}^N \frac{r^{-3}_i}{\sum_{j=1}^Nr^{-3}_j}\boldsymbol{m}_i,
\end{equation}
where $i$ (or $j$) denotes each of the $N$ nearest-neighbor Gd$^{3+}$ ions at the muon site ($N=2$ in our case). Since all Gd moments are parallel to the applied magnetic field, this term results in a constant shift in both components of the Knight shift proportional to the hyperfine coupling $A$. Furthermore, since the two nearest-neighbor Gd atoms are equidistant from the muon site, they contribute equally to the hyperfine field in Eq.~\eqref{eq:hyperfine} and hence the scaling with distance is not important.
The demagnetizing field
\begin{equation}
 \boldsymbol{B}_\mathrm{demag} = -\mu_0 N \boldsymbol{M},   
\end{equation}
where the demagnetizing tensor $N$ depends on the sample shape. For the sample used in this experiment $N$ is not isotropic, and therefore the demagnetizing field exhibits an angular dependence.

We evaluated the demagnetizing tensor for our sample using an analytical solution
obtained for rectangular prisms~\cite{10.1063/1.1703091}. We approximated the sample as a rectangular plate with dimensions 5.7(4)mm by 4.36(2)mm, along $b$ and $c$, respectively, and a thickness of 0.84(2)mm. (The greater uncertainty in the length of the crystal along the $b$ axis reflects the fact that the crystal has two uncut rounded edges). Averaging over the sample volume, this yields a diagonal demagnetizing tensor $N$ with eigenvalues $0.717(2),0.123(5)$ and $0.160(4)$.

The results of our fits to the muon Knight shift at the two muon sites are shown by the red and blue lines in Fig.~\ref{fig:ks}(b) and show good agreement with the data. 
Best fits are obtained with $\alpha=-64(6)^\circ$, $\beta=17(2)^\circ$ and $\gamma=68(3)^\circ$. This set of Euler angles results in an angle of 16$^\circ$ between the sample rotation axis and [010], while the angle between the initial magnetic field direction and [100] is given by $\beta=17(2)^\circ$. Such a degree of misalignment is very plausible, given the experimental setup.
The fitting yields a value $A=-0.0094(4)$~\AA{}$^{-3}$ for the hyperfine coupling constant. The fitted Gd moment is $\mu_\mathrm{Gd}=0.46(2)\mu_\mathrm{B}$. Magnetometry measurements~\cite{Note1} made on this sample at 0.5~T give a moment of $\mu_\mathrm{Gd}=0.437(15)\mu_\mathrm{B}$ at 55 K. Assuming that the susceptibility is independent of field at this temperature would give a moment $\mu_\mathrm{Gd}=0.68(2)\mu_\mathrm{B}$ at 0.78~T, which is slightly larger than the moment suggested by this analysis. 

In summary, the results of these measurements are consistent with muons stopping in GdRu$_2$Si$_2$ at a single crystallographically distinct site. The observation of two distinct muon Knight shifts rules out sites with axial symmetry, such as the site at (0.5, 0.5, 0) that has been proposed for the isostructural compound CeRu$_2$Si$_2$ \cite{amato}, which corresponds to a single magnetically distinct site for all moment directions. As seen in Fig.~\ref{fig:ks}(b), a good fit can be made to the data by assuming that site 1 is the muon stopping site. These measurements therefore provide experimental confirmation for the muon site that we calculated using DFT, and which serves as the basis for much of the analysis in the rest of this paper. They also indicate the presence of a significant negative hyperfine field at the muon site, which will be important when considering the magnetic field at the muon site measured for other phases in GdRu$_2$Si$_2$. Negative hyperfine coupling constants of $-0.45$ and $-0.518$ T$\mu^{-1}_\mathrm{B}$ were found for the paramagnetic phases of MnGe~\cite{PhysRevB.93.174405} and MnSi~\cite{PhysRevB.89.184425}, respectively. These are significantly larger than the value found here for GdRu$_2$Si$_2$, which in these units is $A=-0.073(3)$~T$\mu^{-1}_\mathrm{B}$, but give rise to similar hyperfine fields once the larger moment size of the Gd$^{3+}$ ions is taken into account.

\subsection{Transverse-field $\boldsymbol \mu^+$SR measurements}\label{sec:tf}

Transverse-field $\mu^+$SR measurements were made on aligned crystals of GdRu$_2$Si$_2$ using the HAL-9500 spectrometer at the Swiss Muon Source. In these measurements, the muon spin precesses about the sum of the external magnetic field, which is oriented perpendicular to the initial muon-spin direction, and any internal magnetic fields present in the sample. Two sets of measurements were made: one with the crystals oriented such that applied field $\boldsymbol{B}_0$ pointed along [001] and another set with the crystals oriented such that $\boldsymbol{B}_0$ pointed along the [100] direction. Spectra were measured after zero-field cooling the sample.

\begin{figure}[ht]
	\centering
	\includegraphics[width=\columnwidth]{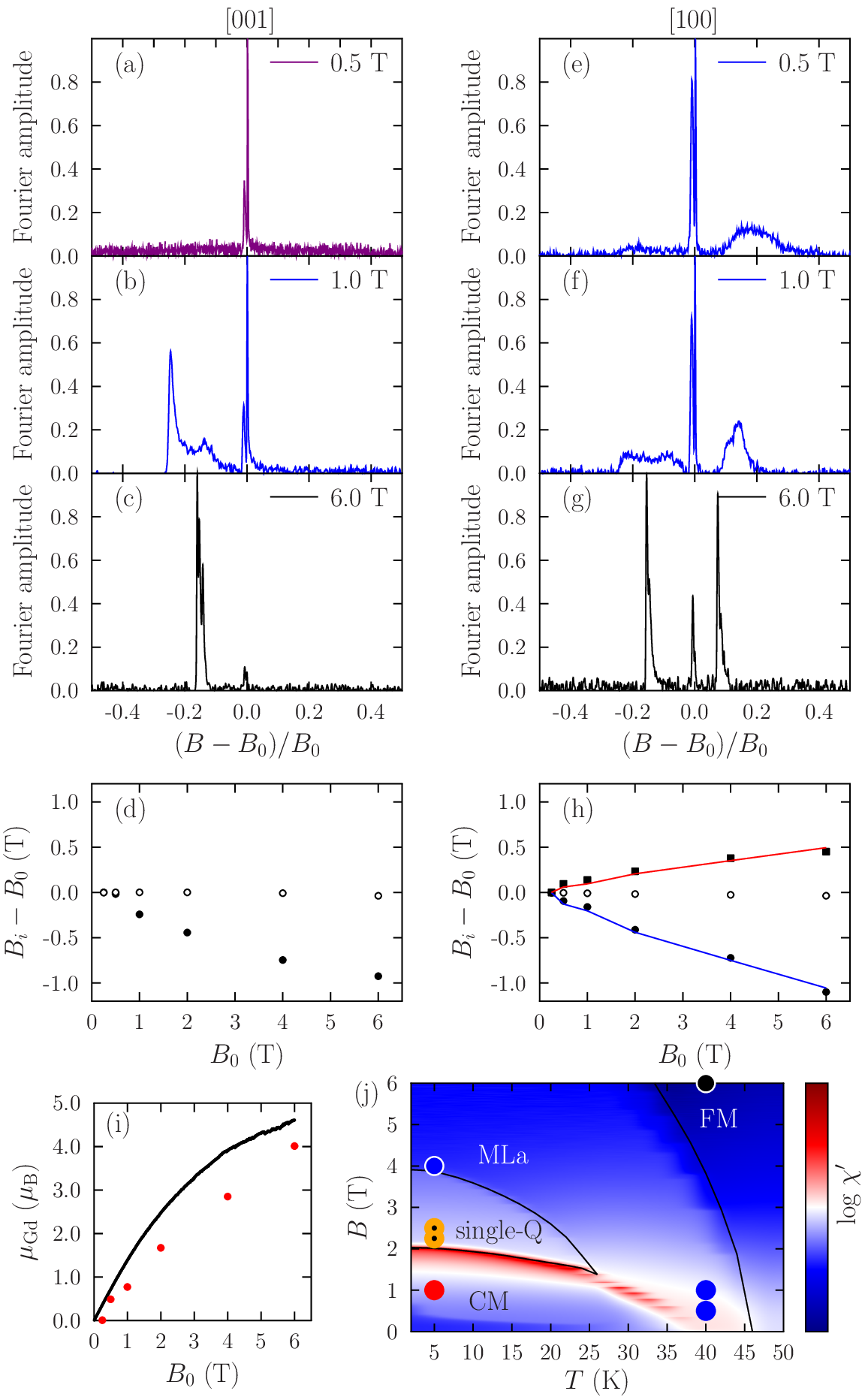}
	\caption{Fourier transform TF $\mu^+$SR spectra measured at $T=40$~K for applied magnetic fields along (a--c) [001] and (e--g) [100].  Positions of  peaks as a function of applied field $B_0$ along [001] and [100] are shown in (d) and (h), respectively. Red and blue lines in (h) correspond to fits of the internal field at muon sites lying along the $b$ axis or $a$ axis respectively, as a function of the extracted Gd magnetic moment.  (i) Magnetic moment of the Gd ions at 40~K, obtained from the separation of the peaks in the  muon spectra (red circles) and magnetometry (black). (j) Real part of the ac susceptibility as a function of temperature and magnetic field for applied magnetic fields along [100]. Circles on the phase diagram correspond to the FT $\mu^+$SR spectra presented in this paper, with  colors corresponding  the lines used to  the spectra.}
	\label{fig:spectra_40K}
\end{figure}

\begin{figure}[ht]
	\centering
	\includegraphics[width=\columnwidth]{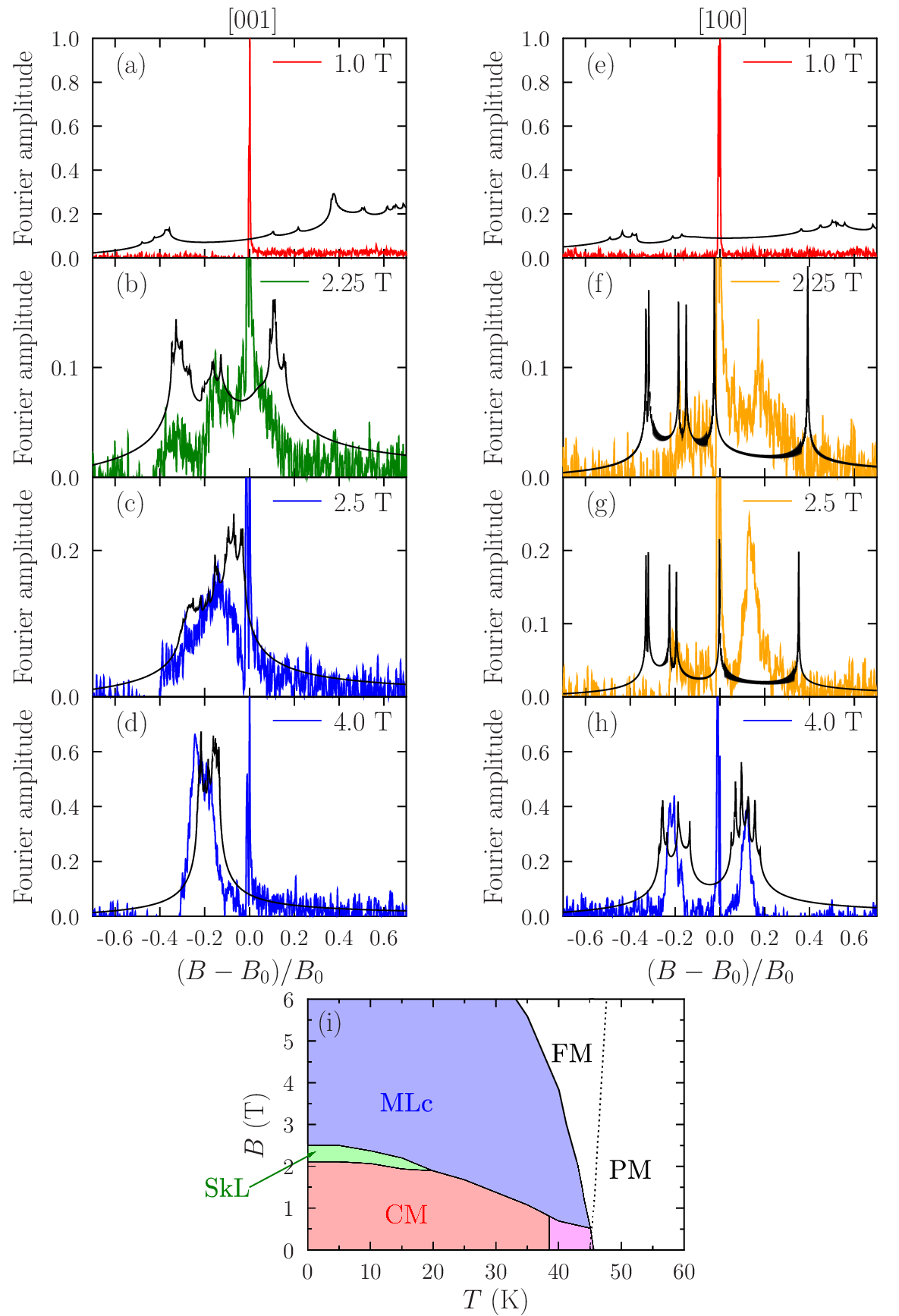}
	\caption{Fourier transform TF $\mu^+$SR spectra measured at $T=5$~K for applied magnetic fields along (a--d) [001] and (e--h) [100]. Black lines indicate simulations of the spectra. (i) Magnetic phase diagram for applied fields along [001], using phase boundaries obtained in Ref.~\cite{Khanh2020}.}
	\label{fig:spectra_5K}
\end{figure}

We first consider the spectra measured at $T=40$~K, where we can access the field-polarized state. Fourier transform (FT) $\mu^+$SR spectra measured at different fields are shown in Fig.~\ref{fig:spectra_40K} for each of the orientations. For  $\boldsymbol{B}_0 \parallel [001]$ the spectra comprise a peak close to the applied field and a secondary peak at lower fields which becomes more prominent as the field is increased. (We note that the central peak is split in two for all of the TF spectra measured on this material, which could result from a fraction of the muons stopping in parts of the sample where the internal field almost, but not completely, cancels, such as, for example, domain walls or surfaces.) For $\boldsymbol{B}_0 \parallel$ [100] there is an additional peak at fields larger than $B_0$ that is absent for the other orientation. We therefore fit the spectra for fields along [001] and [100] to the sum of two or three Lorentzians, respectively, and plot the behavior of the peak centers in Figs.~\ref{fig:spectra_40K}(d) and \ref{fig:spectra_40K}(h). We see that the magnitudes of the field shifts at the two peaks initially increases approximately linearly with the applied field, but then begins to level off, suggesting that the magnetization is starting to saturate. We also note the absence of any satellite peaks for  $B_0 = 0.5$~T $\parallel$ [001] in Fig.~\ref{fig:spectra_40K}(a), with the magnetic phase diagram for this orientation [shown in Fig.~\ref{fig:spectra_5K}(i)] having a phase boundary between 0.5 and 1~T at this temperature. This suggests that the low-field, higher-temperature magnetically ordered state gives rise to a broad distribution of magnetic fields at the muon site, relaxing the contribution of muons stopping in the sample out of the TF spectra, a scenario which is consistent with our ZF spectra, discussed in Sec.~\ref{sec:zf}, which show a large relaxation rate at this temperature. On the other hand, spectra measured at 0.5 and 1.0~T for $\boldsymbol{B}_0 \parallel [100]$, shown in Figs.~\ref{fig:spectra_40K}(e) and \ref{fig:spectra_40K}(f), respectively, are expected to occupy the same magnetic phase [as seen in our ac susceptibility measurements shown in Fig.~\ref{fig:spectra_40K}(j)] and both show satellite peaks. This suggests the zero-field ordered state survives up to higher fields applied along [001] than for those along [100].

The number of peaks observed for each of the field orientations reflects the symmetry of the muon stopping site. The local magnetic field at the muon site is the sum of
the externally applied field and dipolar, Lorentz, demagnetizing and hyperfine fields. For Gd moments along $c$, muons at site 1 will all experience an internal field of $(-0.063 - 0.140N + 7.77A)\mu_\mathrm{Gd}$~T along $c$, where the first term is sum of the dipolar and Lorentz fields, the second term is the demagnetizing field, with the demagnetizing factor $N$ depending on the sample shape, and the final term is the hyperfine field, which is proportional to the hyperfine coupling constant $A$, measured in \AA$^{-3}$. All of these contributions to the internal field are proportional to the Gd magnetic moment $\mu_\mathrm{Gd}$, which in this expression is measured in Bohr magnetons. On the other hand, for moments along $a$, instances of site 1 lying on the $b$ axis will experience a field $(0.295 - 0.140N + 7.77 A)\mu_\mathrm{Gd}$~T along $a$, whereas the crystallographically equivalent site lying on the $a$ axis will instead experience a field $(-0.091 - 0.140N + 7.77 A)\mu_\mathrm{Gd}$~T. For $\boldsymbol{B}_0 \parallel$ [100] the separation of the satellite peaks, $\Delta B$, is therefore related to $\mu_\mathrm{Gd}$ by $\Delta B = 0.386\mu_\mathrm{Gd}$~T. The development of $\mu_\mathrm{Gd}$ as a function of applied magnetic field is shown in Fig.~\ref{fig:spectra_40K}(i). We have also measured the Gd moment as a function of magnetic field using magnetometry \cite{Note1}  [Fig.~\ref{fig:spectra_40K}(i)]. The values of $\mu_\mathrm{Gd}$ obtained using these approaches show reasonable agreement, though we note that those obtained from magnetometry are consistently larger and have a slightly different functional dependence with applied field. Taking the sample geometry to be a flat plate with a perpendicular magnetization, we assume $N \approx 1$. (We note the larger sample area obtained for this mosaic of crystals, compared to a single crystals, means that we expect the out of plane component of the demagnetizing tensor to be closer to 1 than was the case of the sample using in our angle dependent Knight shift measurements).  Fitting either the positive or negative field shifts for $\boldsymbol{B}_0 \parallel$ [100] to a straight line in $\mu_\mathrm{Gd}(B_0)$ [blue or red lines in Fig.~\ref{fig:spectra_40K}(h), respectively] yields $A=-0.0040(9)~\mathrm{\AA}^{-3}$, which is similar to the hyperfine coupling obtained from our analysis of angle-dependent muon Knight shift measurements.  This negative hyperfine contribution can be explained by the spin-polarization of conduction electrons through the 
RKKY interaction, which has been predicted to play an important role in stabilizing the skyrmion lattice in GdRu$_2$Si$_2$ \cite{PhysRevLett.128.157206}.  The multiplicity of the peaks for fields along [100] is consistent with our calculated muon site and with our TF measurements made in the paramagnetic phase at 55~K.

We now turn to the spectra measured at $T=5$~K, where several distinct magnetic structures  exist as a function of applied magnetic field \cite{Khanh2020,https://doi.org/10.1002/advs.202105452}. FT $\mu^+$SR spectra  are shown in Fig.~\ref{fig:spectra_5K} for each orientation. For $B < 2.25$~T, the spectra comprise a split peak,  close to the applied  field.  For $\boldsymbol{B}_0 \parallel$ [001] [Figs.~\ref{fig:spectra_5K}(b--d)], the spectra contain additional spectral weight at fields below the applied field, whereas for $\boldsymbol{B}_0 \parallel$ [100], we see a pair of satellite peaks,  above and below $B_0$. As before, this reflects the symmetry of the muon site for each field orientation. However, the features in the spectra are much broader for $T=5$~K, suggesting that the system adopts magnetic structures that give rise to a wider distribution of  fields. These become sharper at higher applied fields, as the field-polarized component of the magnetic structure becomes more significant.

Knowledge of the muon stopping site and the candidate magnetic structure  allows us to simulate the expected $\mu^+$SR spectra~\cite{Note1}. Simulated spectra for $T=5$~K are shown alongside the data in Fig.~\ref{fig:spectra_5K}. 
(Since the sharp peak at the applied field is due to muon stopping in the silver sample holder, this does not appear in our simulations.) As seen in Fig.~\ref{fig:spectra_40K}(i), the Gd$^{3+}$ moment measured using $\mu^+$SR is consistently smaller than the  7.0$\mu_\mathrm{B}$ expected for an isolated ion, in contrast to the results of two neutron diffraction studies~\cite{PhysRevB.107.L180402,paddison2024spindynamicscentrosymmetricskyrmion}, which have shown that the size of the ordered moment is broadly consistent with 7.0$\mu_\mathrm{B}$. For our simulations, we take $\mu_\mathrm{Gd}=4.8(1)\mu_\mathrm{B}$, deduced from zero-field $\mu^+$SR at low temperatures presented in Sec.~\ref{sec:zf}, where possible reasons for this discrepancy are also discussed.

For the spectra measured at $B_0=1.0$~T [Figs.~\ref{fig:spectra_5K}(a, e)], we see only a sharp peak at the applied magnetic feature, with no other notable features. This peak is due to muons stopping outside of the sample, suggesting that muons stopping inside the sample experience a broad distribution of magnetic fields and therefore have their contribution to the TF spectra dephased. This is consistent with our simulations based on the constant-moment magnetic structure proposed in Ref.~\cite{PhysRevB.107.L180402} [denoted CM in Fig.~\ref{fig:spectra_5K}(i)], which leads to a Fourier spectrum with weight spread across a wide frequency window with no notable peaks. The simulated spectrum for $\boldsymbol{B}_0 \parallel$ [001] does, however, predict some additional spectral weight at fields around 0.5 T above the applied field, which is not seen in experiment, although the overall broadness of the spectrum means that this would likely be difficult to resolve in practice. 

We first focus on the spectra measured for $\boldsymbol{B}_0 \parallel$ [001]. For $B_0=2.25$~T, GdRu$_2$Si$_2$ has been proposed to host a square SkL phase~\cite{Khanh2020}. The corresponding $\mu^+$SR spectrum in Fig.~\ref{fig:spectra_5K}(b) exhibits two peaks at fields below the applied field, whose locations show good agreement with those simulated for an approximate realization of a square SkL. Our simulation also predicts a peak just above the applied field, suggesting that a  fraction of the implanted muons will be at sites where the internal field is well aligned with the external field. Due to the anisotropic nature of the dipolar tensor at the muon site, a magnetic structure with moments polarized along the $c$ axis results in a dipolar field that points in the opposite direction to the magnetization, and hence the external field, as seen in Fig.~\ref{fig:spectra_40K}(c). This therefore highlights the role of the incommensurate components of the magnetic moments in giving rise to a local field at the muon site aligned with the externally-applied field. However, we note that this high-field feature is not resolved in the measured spectrum, possibly due to being too close to the central peak. 

The 2.25~T spectrum is qualitatively different from those measured at 2.5 [Fig.~\ref{fig:spectra_5K}(c)] and 4 T [Fig.~\ref{fig:spectra_5K}(d)] for this orientation, allowing us to clearly distinguish the response of the muon to the SkL from that arising from the other multi-Q magnetic structures that are realized in this system. For the higher fields  along [001], this has been proposed to be the checkerboard antiferroic order of meron/anti-meron-like textures (MLc) \cite{https://doi.org/10.1002/advs.202105452}. The $B_0\geq2.5$~T spectra both comprise a  satellite peak below the applied field, although the shape of the satellite  is distinct in each case. Our simulated spectra based on the MLc magnetic structure show fair agreement with experiment for both values of $B_0$. The difference between the spectra measured at 2.25 [Fig.~\ref{fig:spectra_5K}(b)] and 2.5~T [Fig.~\ref{fig:spectra_5K}(c)] can be explained by the fact that the helical components of the SkL magnetic structure rotate into and out of the $c$-axis direction, either aligning or antialigning with the ferromagnetic component, whereas the modulated parts of the moment in the MLc phase remain within the $ab$ plane (and therefore perpendicular to the ferromagnetic component). Thus the former magnetic structure more effectively splits the spectral weight into distinct peaks, whereas the latter only results in broadening of the low-field feature.

Turning to the spectra measured for $\boldsymbol{B}_0 \parallel$ [100], we note that, in contrast to $\boldsymbol{B}_0 \parallel$ [001], applied fields with magnitudes $B_0=2.25$~T and $B_0=2.5$~T in this orientation are expected to result in a single-Q conical magnetic structure.
The spectra measured in this phase [Fig.~\ref{fig:spectra_5K}(f) and Fig.~\ref{fig:spectra_5K}(g)] comprise pairs of satellite peaks, one above $B_0$ and one below. These features are fairly broad and somewhat difficult to resolve in experiment for $B_0=2.25$~T, but become sharper at $B_0=2.5$~T. Our simulated spectra can be interpreted as a pair of shifted Overhauser distributions~\cite{PhysRevB.89.184425} resulting from a helical magnetic structure propagating along the $a$ axis, which leads to different magnetic fields for muons sitting along the $a$ axis to those lying on the $b$ axis. An additional ferromagnetic component shifts the centers of these distributions, such that one is centered on a field above $B_0$ and one on a field below, and also results in a small splitting of the peaks within this distribution, which is particularly noticeable for the distribution at fields below $B_0$.  It is difficult to reconcile these spectra with those observed in experiment. The peaks observed in the experimental spectra suggest the presence of a significant field-polarized ferromagnetic component, which was included in the magnetic structure used in our simulations, but these are broadened, rather than experiencing the splitting that our simulations predict. We therefore suggest that these spectra are incompatible with the proposed single-Q magnetic structure. It is possible that the magnetic structure for this phase needs to be revised, and further work on this phase would be instructive.

At $B_0=4$~T, an applied field parallel to [100]  gives rise to another checkerboard phase, where the modulations in magnetic moment are rotated relative to those for $\boldsymbol{B}_0 \parallel$ [001], in order to remain perpendicular to the applied field~\cite{https://doi.org/10.1002/advs.202105452}.
For both field orientations, the magnetic structures at $B_0=4$~T contain substantial field-polarized components, so the locations and multiplicity of the peaks mostly reflect the symmetry of the muon sites, as was discussed for the spectra measured at $T=40$~K.

\begin{figure}[htb]
	\centering
	\includegraphics[width=\columnwidth]{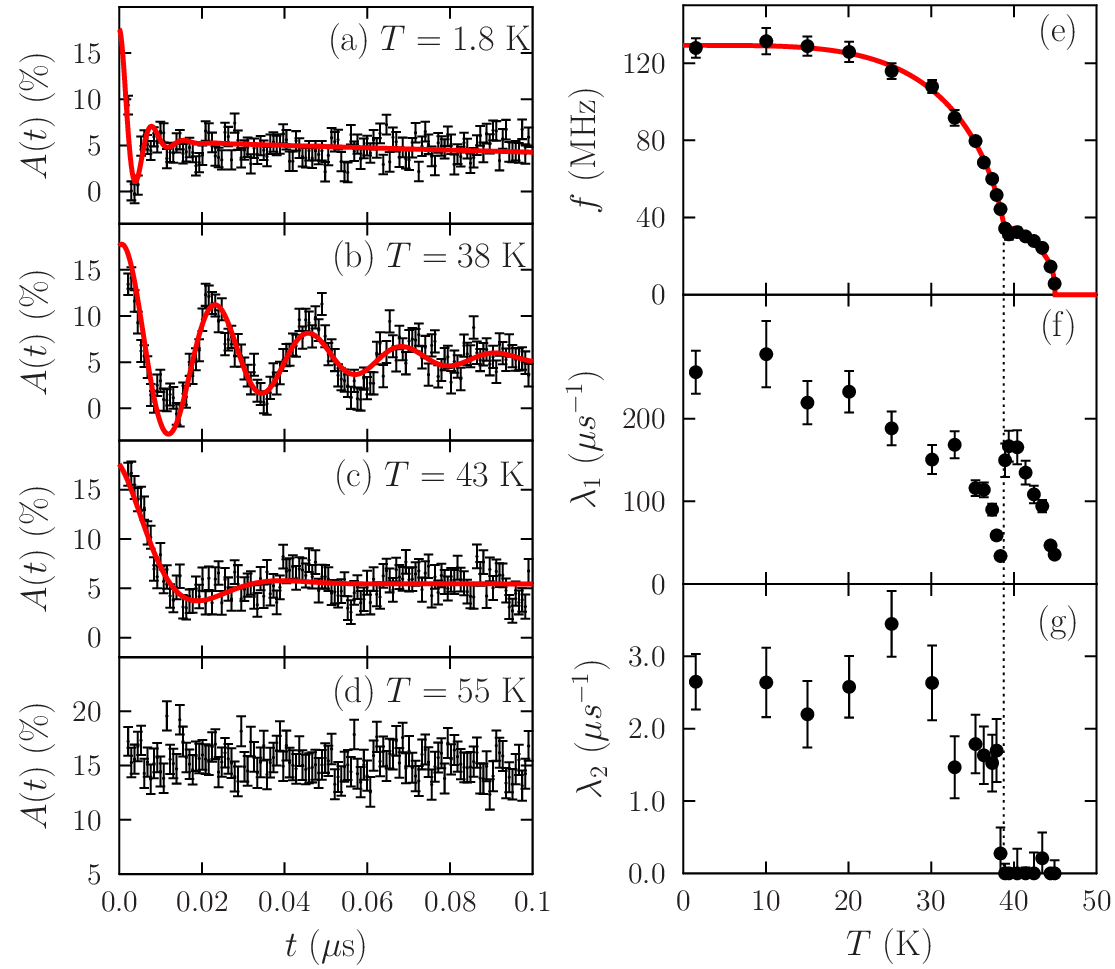}
	\caption{Zero-field $\mu^+$SR spectra measured  at (a) $1.8$~K, (b) $38$~K, (c) $43$~K and (d) $55$~K. Temperature dependence of (e) the muon procession frequency $f$ and the relaxation rates (f) $\lambda_1$ and (g) $\lambda_2$ in Eq.~\eqref{eq1}. 
 }
	\label{fig:zf}
\end{figure}

\subsection{Zero-field $\boldsymbol \mu^+$SR measurements}\label{sec:zf}

Zero-field $\mu^+$SR measurements were made on two aligned GdRu$_2$Si$_2$ single crystals using the GPS spectrometer. The ZF $\mu^+$SR spectra exhibit a very quickly relaxing oscillating asymmetry. The first 0.1 $\mu$s of the spectra were fitted to the function
\begin{equation}\label{eq1}
	A(t)=A_1 \exp(-\lambda_1 t) \cos (2\pi f t + \phi) + A_2 \exp(-\lambda_2 t), 
\end{equation}
where the oscillating and non-oscillating components of the asymmetry have amplitudes $A_1$ and $A_2$ and relaxation rates $\lambda_1$ and $\lambda_2$, respectively. The amplitudes $A_1$ and $A_2$ and the phase $\phi$ were globally refined, with values $A_1 = 12.6(3)\,\%$, $A_2 = 5.56(4)\,\%$ and $\phi = - 17(2)^\circ$. The temperature dependence of the variable parameters is shown in Fig.~\ref{fig:zf}. The muon precession frequency $f$ serves as a magnetic order parameter for the system and exhibits a discontinuity in its derivative that suggests a transition between two distinct magnetic phases, both ordered throughout the bulk.
The frequencies $f(T)$ were fitted to
\begin{equation}
	f(T)=
	\begin{cases}
		f_1 [1-(T/T_\mathrm{c})^\alpha]^{\beta_1} & \text{for } 0 \leq T \leq T_{\mathrm{N}1} \\
		f_2 (1-T/T_{\mathrm{N}2})^{\beta_2} & \text{for }  T_{\mathrm{N}1} \leq T \leq T_{\mathrm{N}2} \\
	\end{cases},
\end{equation}
where $f_2 = \frac{[1-(T_{\mathrm{N}1}/T_\mathrm{c})^\alpha]^{\beta_1}}{(1-T_{\mathrm{N}1}/T_{\mathrm{N}2})^{\beta_2}}$, ensuring that $f(T)$ is continuous at $T=T_{\mathrm{N}1}$. This fitting yields $f_1=129(3)$~MHz, $\alpha=4.06(8)$, $\beta_1=0.49(8)$, $\beta_2=0.36(2)$, $T_\mathrm{c}=39.5(1)$~K, and transition temperatures $T_{\mathrm{N}1}=38.8(1)$~K and
$T_{\mathrm{N}2}=44.95(1)$~K.
The critical exponent $\beta_2 = 0.36(2)$ for the transitions between the higher-temperature ordered phase and the paramagnetic phase shows good agreement with the value $\beta = 0.366$ expected for a 3D Heisenberg magnet \cite{steve_mag}.
The transition at $T_{\mathrm{N}1}$ can also be seen in the relaxation rates, with $\lambda_1$ [Fig.~\ref{fig:zf}(f)] and $\lambda_2$ [Fig.~\ref{fig:zf}(g)] exhibiting discontinuous increases and decreases, respectively.

\begin{figure}[htb]
 	\centering
 	\includegraphics[width=\columnwidth]{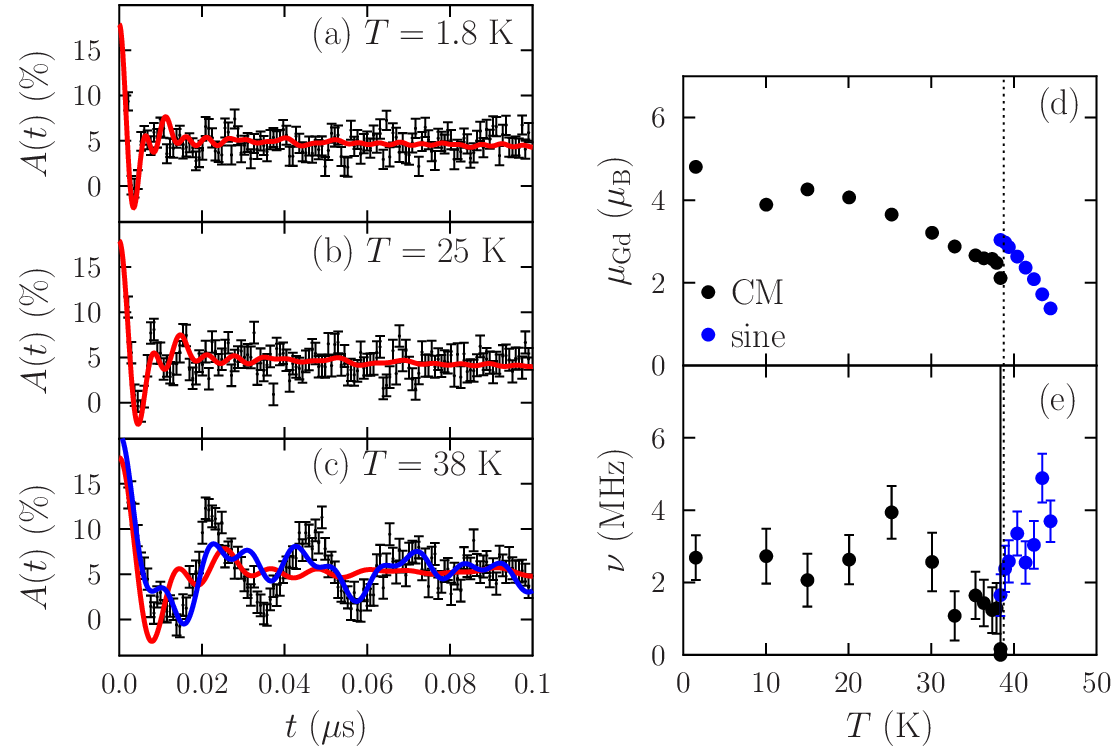}
 	\caption{Fits to a model based on the constant-moment solution for data measured at (a) 1.8 K, (b) 25 K and (c) 38 K. The blue line in (c) indicates a fit to an alternative model based on a sine magnetic structure. (d) Gd magnetic moments and (e) fluctuation rates $\nu$ extracted using simulations of the $\mu^+$SR spectra as a function of temperature.}
 	\label{fig:model_fitting}
 \end{figure}

We have simulated the ZF $\mu^+$SR data across the temperature range, using candidate magnetic structures proposed in other studies~\cite{PhysRevB.107.L180402,paddison2024spindynamicscentrosymmetricskyrmion}. We have modelled the data for $T<T_\mathrm{N1}$ using the double-Q constant-moment (CM) magnetic structure with topological charge stripes proposed in Ref.~\cite{PhysRevB.107.L180402}. Fitting the simulated spectra to the data provides the $T$ dependence of the Gd magnetic moment $\mu_\mathrm{Gd}$. Furthermore, the effect of dynamics has been incorporated in these simulations using the strong collision approximation, thereby providing an estimate for the characteristic rate $\nu$ of spin fluctuations. Fits at representative temperatures, as well as the temperature-dependence of these parameters are shown in Fig.~\ref{fig:model_fitting}. We find an ordered moment of 4.8(1)$\mu_\mathrm{B}$ at $T=1.5$~K, which is $\approx$30\% smaller than the full ionic value of 7.0$\mu_\mathrm{B}$. The Gd moment is obtained from the scale factor relating the simulated magnetic fields at the muon site to those observed in experiment, and hence this discrepancy could result from contributions to the magnetic field that are not accounted for by our model, such as those from the itinerant electrons.

The extracted fluctuation rates $\nu\approx$ 3~MHz are significantly smaller than the relaxation rates $\lambda_1$ that accompany the relaxing part of the asymmetry, indicating that this relaxation is mostly due to the distribution of static magnetic fields resulting from the incommensurate magnetic structure. They are also much smaller than the precession frequency $f$, implying we are in a slowly fluctuating regime (assuming the fluctuating amplitude is of a similar order of magnitude to the static field). The discontinuous behavior of $\lambda_1$ at $T_\mathrm{N1}$ is therefore evidence for a significant change in the magnetic structure at this temperature. The fluctuation rates $\nu$ obtained from our simulations are similar in magnitude to the slow relaxation rate $\lambda_2$, consistent with the interpretation that this relaxation is due to dynamics flipping the muon spin at sites where the local magnetic field is parallel to the muon spin, in the slow fluctuation limit. We note that the relative magnitudes of $A_1$ and $A_2$ are consistent with those of the oscillating and non-oscillating parts, respectively, of the polarization function obtained by our model.

Unlike the magnetic ground state~\cite{PhysRevB.107.L180402}, the higher-temperature magnetically ordered phase has not been unambiguously identified. However, a recent powder neutron powder diffraction study~\cite{paddison2024spindynamicscentrosymmetricskyrmion} proposed three candidate single-Q magnetic structures for this phase, with these measurements being unable to distinguish between these. We therefore simulated $\mu^+$SR spectra based on each of the models and fit these to the data measured at $T \ge 38$~K. All of these models provided similar quality fits to the data, with a so-called sine magnetic structure with moments along the [011] direction being the best, and also being the only model that gives a magnetic moment that decreases monotonically with temperature, as shown in Fig.~\ref{fig:model_fitting}(d). (We note that for these sine magnetic structures, which do not have a constant magnetic moment amplitude, $\mu_\mathrm{Gd}$ represents the maximum value of the magnetic moment.) This magnetic structure also provides a better fit to the spectrum measured at $T=38$~K than the constant-moment solution, as indicated by the blue line in Fig.~\ref{fig:model_fitting}(c). Our modeling also indicates that the fluctuation rate increases on approaching $T_\mathrm{N2}$ from below, as shown in Fig.~\ref{fig:model_fitting}(e), but remains within the slow fluctuation limit. This suggests that the distribution of static magnetic fields resulting from an incommensurate magnetic structure is also responsible for the majority of the relaxation observed in this higher-temperature ordered phase. 

\section{Conclusion}\label{sec:conclusion}
We have illustrated how implanted muons are sensitive to the magnetic phases in GdRu$_2$Si$_2$, providing evidence for skyrmion and meron-lattice magnetic phases. In contrast, $\mu^+$SR spectra measured for fields along [100] are inconsistent with the previously proposed single-Q magnetic structure, suggesting that this phase needs to be revised. We demonstrate a significant, negative hyperfine contribution to the local magnetic field at the muon site, due to conduction electrons becoming spin-polarized through the RKKY interaction. This significant hyperfine field is also observed in our angle-dependent measurements of the muon Knight shift, which also provide strong experimental evidence for the muon stopping sites that form the basis of much of our analysis. Zero field measurements as a function of temperature demonstrate a transition between two distinct bulk magnetically ordered states at $T_{\mathrm{N}1} = 38.8(1)$~K.

Data presented in this paper will be made available via Ref.~\cite{data}.

\begin{acknowledgments}
Part of the work was carried out at the Swiss Muon
Source, Paul Scherrer Institute, Switzerland and we are grateful for the provision of beamtime and experimental support.
We acknowledge computing resources provided by Durham Hamilton HPC. This work was funded by the EPSRC UK Skyrmion Project under Grant No. EP/N032128/1. The work at the University of Warwick was further supported by EPSRC grant EP/T005963/1. Work at Oxford was funded by UK Research and Innovation (UKRI) under the UK government’s Horizon Europe funding guarantee (Grant No. EP/X025861/1). AHM is grateful to STFC and EPSRC for the provision of a studentship under Grant No. EP/T518001. MG acknowledges the financial support of the Slovenian Research and Innovation Agency through Program No. P1-0125 and Projects No. Z1-1852, N1-0148, J1-2461, J1-50008, J1-50012, N1-0356, and N1-0345. 
\end{acknowledgments}

\end{document}


\title{Supplemental Material for ``Field-orientation-dependent magnetic phases in GdRu$_2$Si$_2$ probed with muon-spin spectroscopy''}
\author{B.\,M. Huddart}
\affiliation{Department of Physics, Durham University, Durham DH1 3LE, United Kingdom}
\affiliation{Clarendon Laboratory, University of Oxford, Department of Physics, Oxford OX1 3PU, United Kingdom}
\author{A. Hern\'{a}ndez-Meli\'{a}n}
\affiliation{Department of Physics, Durham University, Durham DH1 3LE, United Kingdom}
\author{G.\,D.\,A. Wood}
\affiliation{University of Warwick, Department of Physics, Coventry CV4 7AL, United Kingdom}
\author{D.\,A. Mayoh}
\affiliation{University of Warwick, Department of Physics, Coventry CV4 7AL, United Kingdom}
\author{M. Gomil\v{s}ek}
\affiliation{Jo\v{z}ef Stefan Institute, Jamova c. 39, SI-1000 Llubljana, Slovenia}
\affiliation{Faculty of Mathematics and Physics, University of Ljubljana, Jadranska u. 19, SI-1000 Llubljana, Slovenia}
\author{Z. Guguchia}
\affiliation{PSI Center for Neutron and Muon Sciences CNM, 5232 Villigen PSI, Switzerland}
\author{C. Wang}
\affiliation{PSI Center for Neutron and Muon Sciences CNM, 5232 Villigen PSI, Switzerland}
\author{T. J. Hicken}
\affiliation{PSI Center for Neutron and Muon Sciences CNM, 5232 Villigen PSI, Switzerland}
\author{S.\,J.\,Blundell}
\affiliation{Clarendon Laboratory, University of Oxford, Department of Physics, Oxford OX1 3PU, United Kingdom}
\author{G. Balakrishnan}
\affiliation{University of Warwick, Department of Physics, Coventry CV4 7AL, United Kingdom}
\author{T. Lancaster}
\affiliation{Department of Physics, Durham University, Durham DH1 3LE, United Kingdom}
\date{\today}

\maketitle

\section{Muon-spin relaxation}
In a $\mu^{+}$SR experiment \cite{muon_book} spin polarized muons
are implanted into the sample, where they precess about the total magnetic field $B$ at the muon site at a frequency $\nu=\gamma_{\mu} B / 2\pi$, where $\gamma_{\mu}$ is the muon gyromagnetic ratio ($=2\pi \times 135.5~\text{MHz}~\text{ T}^{-1}$).  These muons decay with an average lifetime of 2.2~$\mu$s into two neutrinos and a positron.  Due to the parity-violating nature of the weak interaction, positrons are emitted preferentially along the instantaneous muon spin direction.  Recording the direction of emitted positron therefore allows us to infer the muon spin polarization at the time of decay.  The quantity of interest is the asymmetry 
\begin{equation}
	A(t)=\frac{N_{\textrm{B}}(t)-\alpha N_{\textrm{F}}(t)}{N_{\textrm{B}}(t)+\alpha N_{\textrm{F}}(t)},
	\label{eq:zf_asymmetry}
\end{equation}
where $N_{\textrm{F/B}}$ is the number of positrons detected in the forward/backward detectors and $\alpha$ is an experimental calibration constant.  The asymmetry $A(t)$ is proportional to the spin polarization of the muon ensemble.
For a magnetically-ordered compound one observes oscillations in the asymmetry $A(t)$.  For a distribution of magnetic fields the spins will each precess at a different frequency, resulting in a relaxation of the muon polarization. 

Zero-field $\mu^+$SR measurements were carried out using the GPS spectrometer at the Swiss Muon Source, Paul Scherrer Institute (Switzerland). Two single crystal samples, in the form of platelets with masses 124 and 127 mg, were enclosed in Kapton tape and mounted on a sample holder (fork) with their large (001) faces perpendicular to the beam. These measurements were made in a spin-rotated configuration, with the initial muon-spin polarization lying at an angle $\approx 45^\circ$ with the muon beam axis. Hence, the polarization along the initial muon spin direction was found by forming the appropriate pairs of detectors (forward and up; and backward and down). Similarly, our simulations of the $\mu^+$SR spectra consider the precession of components of the muon-spin polarization along each of these detector directions.

Transverse-field measurements were carried out using the HAL-9500 spectrometer at the Swiss Muon Source. Two sets of measurements were made. For the measurements with the applied field along [001] we used three single crystal samples with masses 119, 124 and 127~mg. For the measurements with the applied field $\boldsymbol{B_0} \parallel [100]$ we used two different crystals with masses 149 and 225 mg. In each case the samples were secured on a silver sample holder using GE varnish, with their large faces perpendicular to the incoming beam. Spectra were measured after zero-field cooling the sample.

Data analysis was carried out using the WiMDA program \cite{wimda}, with TF spectra generated using WiMDA’s apodized, phase-corrected cosine Fourier transforms. A Gaussian filter was applied to the data before carrying out the Fourier transform. Due to the wide range of fields spanned by our measurements, we used a different filter time constant in each case. A shorter time constant reduces the noise in the spectrum (due to removing later parts of the spectrum in the time domain where the asymmetry has relaxed to a very small amplitude) at the expense of being able to resolve fine details in the spectra, which may or may not be physical. We found values for the time constant $\tau(B_0)=200/(\gamma_{\mu} B_0$), where $B_0$ is the magnitude of the applied magnetic field, to provide a good compromise. This represents the time taken for 200 oscillations of the asymmetry at the Larmor frequency of the applied field in each case.  We also made use of the MINUIT algorithm \cite{minuit} via
the iminuit \cite{iminuit} Python interface for global refinement of fit parameters.

\section{Muon stopping sites}
We have carried out density functional theory (DFT) calculations using the plane-wave basis-set electronic structure code \textsc{castep} \cite{CASTEP} in order to determine the muon stopping sites. Calculations were carried out within the generalized-gradient approximation (GGA) using the PBE functional \cite{PBE}. GdRu$_2$Si$_2$ crystallizes in the tetragonal $I4/mmm$ (No. 139) space group, with $a=b=4.1634(4)$~\AA{} and $c=9.6102(49)$~\AA{} \cite{HIEBL1983287}. Structural relaxations with the muon were carried out on a supercell comprising $2 \times 2\times1$ conventional unit cells of GdRu$_2$Si$_2$ to suppress the unphysical interaction of the muon with its periodic images in plane-wave DFT.  We used a plane-wave cutoff energy of $1000$ eV and $5 \times 5 \times 5$ Monkhorst-Pack grid \cite{mpgrid1976} for Brillouin zone sampling, resulting in total energies that converge to within 1 meV per atom. The system was treated as spin-polarized in these calculations, with a ferromagnetic arrangement of Gd spins. These parameters resulted in DFT-optimized lattice constants that were within 1\% of experiment; these were therefore fixed at their experimental values for subsequent calculations.

Muon site calculations were carried out using the MuFinder program \cite{mufinder}. Initial structures comprising a muon (modelled as a proton) and the GdRu$_2$Si$_2$ supercell were generated by requiring the muon to be at least 0.5~\AA~away from each of the initial muon positions in the previously generated structures (including their symmetry equivalent positions) and at least 1.0~\AA~away from any of the atoms in the cell.  This resulted in 30 structures which were subsequently allowed to relax until the calculated forces on the atoms were all $<5\times 10^{-2}$ eV \AA$^{-1}$ and the total energy
and atomic positions converged to within $2\times10^{-5}$ eV per atom and $1\times10^{-3}$~\AA, respectively. These structural relaxations result in two distinct muon stopping sites, whose properties are summarized in Table~\ref{tab:sites}. The hyperfine coupling at the muon site was calculated using the Gauge-Including Projector Augmented Wave (GIPAW) method implemented in \textsc{castep}~\cite{magres_review}. These hyperfine coupling constants are similar in magnitude to the hyperfine fields obtained by fitting the experimental data, but have the opposite sign. However, we note that our calculations do not include the effect of a transferred hyperfine coupling due to the RKKY interaction.

\begin{table}[htb]
	\caption{\label{tab:sites}%
		Properties of the two crystallographically distinct muon stopping sites in GdRu$_2$Si$_2$. Muon positions are given in fractional coordinates for the conventional unit cell. Site energies are given relative to that of the lowest-energy site.
	}
	\begin{ruledtabular}
		\begin{tabular}{lcccr}
			\textrm{Site no.}&
			\textrm{Muon position}&
			\textrm{Energy (eV)}&
			$A$ (MHz)&
            $A$ (\AA{}$^{-3}$)\\
			\colrule
			1 & (0.50, 0.00, 0.07) & 0 & 21.7 & 0.003 \\
			2 & (0.21, 0.21, 0.16)& 0.26 & 8.06 & 0.001 \\
		\end{tabular}
	\end{ruledtabular}
\end{table}

We investigated the energy barriers in moving between sites 1 and 2, and also the barrier involved in moving between adjacent instances of site 1 (sites 1 and $1'$). We first determined the intermediate states through transition state searches (TSSs) using a generalized synchronous transit method \cite{govind}. Once the transition states were identified, these were confirmed using the nudged elastic band (NEB) method \cite{NEB}, which also yielded the energy profile presented in Fig. 2(b) in the main text. Note that the increased computational cost associated with the NEB calculations means that we used a $3 \times 3 \times 3$ Monkhorst-Pack grid for Brillouin zone sampling in this case instead. 

Finally, we computed the zero-point energy (ZPE) of the muon at each site, to compare this with the heights of the energy barriers. The light muon mass means that, within the harmonic approximation, the ZPE of the muon can be approximated as the sum of the three highest frequency phonon modes calculated for each of the stopping sites, as long as the muon is in the weakly-bound adiabatic limit (i.e., with zero entanglement with nearby nuclear positions)~\cite{gomilsek2023many}. Phonons were calculated at the $\Gamma$ point for sites 1 and 2 using the finite displacement method. As was the case for the NEB calculations, we used a $3 \times 3 \times 3$ grid for $k$-point sampling, reflecting the significant computational cost of these calculations. To check whether the muon is, in fact, in the weakly-bound adiabatic limit, we can calculate the entanglement witness $w_1$, obtained by projecting the three highest-energy (normalized) phonon normal modes onto pure muon motion, summing the squared norms of the resulting (projected) phonon eigenvectors, and subtracting the value 3~\cite{gomilsek2023many}. This results in $w=0$ for zero muon–nuclear entanglement (weakly-bound adiabatic limit) and $w_1 < 0$ in the entangled, many-body case. We obtained entanglement witnesses $w_1=-0.0079$ and $w_1=-0.0082$ for site 1 and site 2, respectively, which are both close to zero, confirming the validity of the weakly-bound adiabatic approximation, i.e., the decoupling of muon vibrational modes from the vibrational modes of the lattice. Under this approximation, we obtain zero-point energies of 0.56 and 0.65 eV, for sites 1 and 2, respectively.

When assessing whether the ZPE of the muon is sufficient to clear the energy barriers between sites, we want to consider only the part of the ZPE due to the effectively 1D motion along the minimum-energy path. Considering only the three highest-energy phonon modes, the muon's dynamical matrix $\mathcal{D}(\textbf{q})$ can be written \cite{srivastava,kantorovich}

\begin{equation}
	\mathcal{D}(\textbf{q}) = U(\textbf{q})\Omega^2(\textbf{q})U^\dagger(\textbf{q}),
\end{equation}
where $\Omega(\textbf{q})$ is a $3 \times 3$ diagonal matrix with the top three real phonon frequencies $\omega_{(\textbf{q},j)}$ along its diagonal, with $j$ denoting the phonon branch index at this $\textbf{q}$. The matrix $U(\textbf{q})$ is the $3 \times 3$ submatrix of the full phonon unitary matrix (whose columns are given by the normalized polarisation vectors corresponding to each of these phonon branches) obtained by keeping only the rows and columns corresponding to muon displacements. $U(\mathbf{q})$ is thus not necessarily unitary, if muon motion becomes correlated (entangled) with the motion of nuclei. This is precisely what is measured by the muon--nuclear entanglement witness $w_1$~\cite{gomilsek2023many}, which can, in our notation, be written succinctly as: $w_1 = \tr\left( U(0) U(0)^\dagger \right) - 3$, i.e., the deviation of $U(0)$ from unitarity. If $w_1 = 0$ muon motion is decoupled (separable) from the motion of the lattice, while if $w_1 < 0$ muon motion is entangled with the motion of the nuclei in the crystal lattice~\cite{gomilsek2023many}. Note that here we consider only $\textbf{q}=0$, since the muon is a localized defect and thus its vibrational modes have no $\mathbf{q}$-space dispersion. The relevant frequency $\omega_{\boldsymbol{v}}$ for the effective one-dimensional  (1D) harmonic oscillator constrained to allow the muon to only move along a specific normalized tunneling direction $\boldsymbol{v}$ is then given by

\begin{equation}
	\omega^2_{\boldsymbol{v}} = \boldsymbol{v}^T 	\mathcal{D}(0) \boldsymbol{v}.
\end{equation}

For a muon at site 2, which corresponds to the transition state for the path between sites 1 and $1'$, the tangent to the minimum energy path must be parallel to the line joining 1 and $1'$, by symmetry. Hence we have $\boldsymbol{v}=(-1, 1, 0)/\sqrt{2}$ along this tunneling path, in Cartesian coordinates aligned with the crystal axes. This yields an oscillator frequency corresponding to a 1D tunneling ZPE of $\hbar \omega_{\boldsymbol{v}}/ 2 =0.25$~eV. For a muon at site 1, we approximate the tangent to the minimum energy path as the displacement of the muon between site 1 and the next point along the minimum energy path obtained from the NEB calculation. Doing so gives a 1D tunneling ZPE of 0.18 eV.

While a muon at site 1 does not have sufficient energy to climb the barrier between it and site $1'$, quantum tunneling between these sites remains a possibility. The WKB approximation allows us to estimate the transmission coefficient as \cite{griffiths}
\begin{equation}
	\mathcal{T} = \exp\left(-\frac{2}{\hbar} \int_{x_1}^{x_2} \sqrt{2m\left(V(x)-E\right)} \, \mathrm{d}x\right),
\end{equation}
for a one dimensional potential $V(x)$ experienced by a muon with mass $m$ and zero-point energy $E$. Integration is carried out between the classical turning points, i.e., the points where $V(x) = E$. We obtain a transmission probability $\mathcal{T} \approx 1.3 \times 10^{-16}$ for tunneling between site 1 and $1'$. Taking the attempt frequency to be equal to that of the harmonic oscillator frequency for a muon at site 1, i.e, $f = E / h$, we estimate a tunneling rate $f\mathcal{T} \approx 0.05$~Hz. This corresponds to a significantly longer timescale than that over which our measurements are made.

\section{Dipolar field calculations}

\subsection{Magnetic structures}

GdRu$_2$Si$_2$ has been proposed \cite{Khanh2020,https://doi.org/10.1002/advs.202105452,PhysRevB.107.L180402} to host a series of double-Q magnetic structures defined in terms of the propagation vectors $\boldsymbol{q}_1 = (q_1, 0, 0)$ and $\boldsymbol{q}_2 = (0, q_2, 0)$. Using the propagation vector formalism, these structures can be written in the general form
\begin{equation}
	\boldsymbol{m}_\mathrm{ic}(\boldsymbol{r})=\Re \left[\sum_{n,p} \boldsymbol{S}_{np} \exp\left[-2\pi i\left(n\boldsymbol{q}_1 + p\boldsymbol{q}_2 \right) \cdot\boldsymbol{r}\right]\right],
\end{equation}
where $n$ and $p$ are integers. The various double-Q magnetic structures are distinguished by their Fourier components $\boldsymbol{S}_{np}$. In some cases the above expression yields a magnetic structure with a moment whose magnitude varies in space. However, the fact that the saturation magnetization of GdRu$_2$Si$_2$ is 7$\mu_\mathrm{B}$~ \cite{GARNIER199680,Khanh2020}, strongly suggests that the magnetism is due to localized $f$-electron moments of Gd$^{3+}$, whose magnitude would be expected to be constant in space. Thus in some cases it will be necessary to normalize the magnetic structure, viz $\boldsymbol{m}(\boldsymbol{r})=\boldsymbol{m}_\mathrm{ic}/|\boldsymbol{m}_\mathrm{ic}|$. A prescription for carrying out this normalization while remaining within the propagation vector formalism is described in Sec.~\ref{sec:normalization}.

Once the decomposition of the target magnetic structure into its Fourier components is found, we can use the fact that the dipolar field is linear in the magnetic moment to compute the dipolar fields arising from each of the constituent helices within the expansion. The calculations are carried out using the MuESR Python library \cite{doi:10.7566/JPSCP.21.011052}, which makes use of the method described in Ref.~\cite{PhysRevB.93.174405} to efficiently calculate the distribution of magnetic fields due to a helical magnetic structure. Corrected to the dipolar fields resulting from the muon-induced displacement of nearby magnetic ions are accounted for by using the approach described in Ref.~\cite{mufinder}, extended for use with incommensurate magnetic structures. We use a value $A=-0.004~\mathrm{\AA}^{-3}$ for the contact hyperfine coupling constant, determined from an analysis of transverse-field $\mu^+$SR data measured in the field-polarized state, described in the main text. For all the dipolar field calculations presenting in this paper, we use a value $N = 1$ for the demagnetizing factor, corresponding to a sample geometry of an infinite plate with magnetization perpendicular to the plane.

\subsubsection{Constant moment solution}
A constant-moment solution has been proposed as the zero-field magnetic ground state of GdRu$_2$Si$_2$~\cite{PhysRevB.107.L180402}. This construction comprises a helix, a spin-density wave (SDW), and an additional helix, coplanar with the first, but propagating along $\boldsymbol{q}_1+2\boldsymbol{q}_2$. The magnetic moment at position $\boldsymbol{r}$ is then given by
\begin{equation}
	\boldsymbol{m}(\boldsymbol{r})=\begin{bmatrix} 0 \\ m_1 \cos(k_1) \\ m_1 \sin(k_1) \end{bmatrix}+\begin{bmatrix} m_2 \sin(k_2) \\ 0 \\ 0 \end{bmatrix}+\begin{bmatrix} 0 \\ m_3 \cos(k_1+2k_2) \\ m_3 \sin(k_1+2k_2) \end{bmatrix},
\end{equation}
where $k_i=-2\pi\boldsymbol{q}_i \cdot \boldsymbol{r}$. The scalar amplitudes $m_1 = 6.49\mu_\mathrm{B}$, $m_2 = 5.44\mu_\mathrm{B}$ and $m_3 = 1.14\mu_\mathrm{B}$ satisfy the constant-moment condition. The principal propagation vectors have magnitudes $q_1=0.22540(2)$ and $q_2=0.22014(3)$. As the final step of this decomposition, we note that an SDW can be expressed as the sum of two helices of opposite chirality and phase, i.e,
\begin{equation}
	\begin{bmatrix} m_2 \sin(k_2) \\ 0 \\ 0 \end{bmatrix} = \begin{bmatrix} 0.5m_2 \sin(k_2) \\ 0 \\ 0.5m_2 \cos(k_2) \end{bmatrix} + \begin{bmatrix} -0.5m_2 \sin(-k_2) \\ 0 \\ -0.5m_2 \cos(-k_2) \end{bmatrix}.
\end{equation} 
Similar decompositions are employed for the other magnetic structures that comprise an SDW as one of their components. 

\subsubsection{Skyrmion lattice}
The square skyrmion lattice spin texture is described approximately by $\boldsymbol{m}(\boldsymbol{r})=\boldsymbol{m}_\mathrm{sk}/|\boldsymbol{m}_\mathrm{sk}|$, where

\begin{equation}
	\boldsymbol{m}_\mathrm{sk}(\boldsymbol{r})=\begin{bmatrix} 0 \\ -m_q \sin(k_1) \\ m_q \cos(k_1)  \end{bmatrix}+\begin{bmatrix} m_q \sin(k_2) \\ 0 \\ m_q \cos(k_2) \end{bmatrix}+\begin{bmatrix} 0 \\ 0 \\ m_z \end{bmatrix},
\end{equation}
and the ferromagnetic component $m_z$ approximately scales with the field-induced out-of-plane uniform magnetization \cite{Khanh2020}. As described in Sec.~\ref{sec:normalization}, the normalization procedure introduces higher harmonic terms in $\boldsymbol{m}(\boldsymbol{r})$. 

\subsubsection{Checkerboard phases}
We also make use of phases proposed in Ref.~\cite{https://doi.org/10.1002/advs.202105452}, that can be considered as the antiferroic order of meron/anti-meron-like into a checkerboard pattern. For $\boldsymbol{H} \parallel [001]$ we have a phase comprising the sum of two SDWs along the $a$ and $b$ axes, and a ferromagnetic component along $c$,
\begin{equation}
	\boldsymbol{m}(\boldsymbol{r})=\begin{bmatrix} 0 \\ m_b \sin(k_1) \\ 0  \end{bmatrix}+\begin{bmatrix} m_a \sin(k_2) \\ 0 \\ 0 \end{bmatrix}+\begin{bmatrix} 0 \\ 0 \\ m_c \end{bmatrix}.
\end{equation}
We denote this phase (phase III in Ref.~\cite{https://doi.org/10.1002/advs.202105452}) as MLc. For $\boldsymbol{B_0} \parallel [100]$ the moments in the SDW propagating along $\boldsymbol{q}_2$ are instead parallel to $c$ and the uniform magnetization component is along $a$. We therefore have a magnetic structure
\begin{equation}
	\boldsymbol{m}(\boldsymbol{r})=\begin{bmatrix} 0 \\ m_b \sin(k_1) \\ 0  \end{bmatrix}+\begin{bmatrix}  0 \\ 0 \\ m_c \sin(k_2) \end{bmatrix}+\begin{bmatrix} m_a \\ 0 \\ 0 \end{bmatrix}.
\end{equation}
We denote this phase (phase III$^\prime$ in Ref.~\cite{https://doi.org/10.1002/advs.202105452}) as MLa. For both of these magnetic structures we assume that the SDWs propagating along $\boldsymbol{q}_1$ and $\boldsymbol{q}_2$ have equal magnitudes, i.e., $m_a=m_b$ for phase MLc and $m_b=m_c$ for the MLa phase. For the MLa phase, we use the magnetization as a function of applied field $H$ (see Sec.~\ref{sec:magnetometry}) to determine the relative amplitude of the ferromagnetic component. Both of these expression lead to a spin texture with a varying spin density, so must be normalized by following the procedure in Sec.~\ref{sec:normalization}.

\subsubsection{Single-Q state}
The final low-temperature magnetic structure that we consider is a single-Q magnetic state (phase IV in Ref~\cite{https://doi.org/10.1002/advs.202105452}), that is realized in a region of the phase diagram for $\boldsymbol{B_0} \parallel [100]$. This can be straightforwardly written as the sum of a helix propagating along $\boldsymbol{q}_1$ and a uniform magnetization along $a$.
\begin{equation}
	\boldsymbol{m}(\boldsymbol{r})=\begin{bmatrix} 0 \\ m_q \cos(k_1) \\ m_q \sin(k_1) \end{bmatrix}+\begin{bmatrix} m_a \\ 0 \\ 0 \end{bmatrix}.
\end{equation}
The amplitudes $m_a$ and $m_q$ are set by the uniform magnetization, which scales as $m_a$ and the condition that the total Gd moment should be constant as a function of applied field.

\subsubsection{Higher-temperature single-Q phases}

Recently~\cite{paddison2024spindynamicscentrosymmetricskyrmion}, three candidate magnetic structures were put forward for the higher-temperature magnetically ordered phase. These are single-q magnetic structures with $\boldsymbol q = (0.2242,0,0)$. Two of these are so-called sine magnetic structures, which have the form of spin-density waves. The first of these, which we refer to as sine-I, has magnetic moments with components along all of the crystallographic axes and can be written

\begin{equation}
	\boldsymbol{m}(\boldsymbol{r})=\begin{bmatrix} m_a \sin(k) \\ m_b \sin(k) \\ m_c \sin(k) \end{bmatrix},
\end{equation}
where $k=-2\pi \boldsymbol{q} \cdot \boldsymbol{r}$ and the amplitudes are $m_a=1.00\mu_\mathrm{B}$, $m_b=2.17\mu_\mathrm{B}$ and $m_c=2.39\mu_\mathrm{B}$. The small relative magnitude of $m_a$ suggests that a suitable single-parameter solution may be obtained by setting $m_a = 0$ and $m_a = m_c = m_q$. This yielded a magnetic structure, which we denote sine-II, given by 

\begin{equation}
	\boldsymbol{m}(\boldsymbol{r})=\begin{bmatrix} 0 \\ m_q \sin(k) \\ m_q \sin(k) \end{bmatrix},
\end{equation}
where $m_q=2.32\mu_\mathrm{B}$. Note that both of these magnetic structures have moments whose magnitudes vary in space. In powder neutron diffraction~\cite{paddison2024spindynamicscentrosymmetricskyrmion}, this second magnetic structure cannot be distinguished from a helical structure given by
\begin{equation}
	\boldsymbol{m}(\boldsymbol{r})=\begin{bmatrix} 0 \\ m_q \cos(k) \\ m_q \sin(k) \end{bmatrix},
\end{equation}
where $m_q=2.32\mu_\mathrm{B}$, as before.

\subsubsection{Normalization}\label{sec:normalization}

Several of the propagation vectors expansions for the magnetic structures in this section do not result in magnetic moments with a constant magnitude. It is therefore necessary to normalize these magnetic structures at each point in space, i.e, $\boldsymbol{m}(\boldsymbol{r})=\boldsymbol{m}_\mathrm{ic}(\mathbf{r})/|\boldsymbol{m}_\mathrm{ic}(\mathbf{r})|$. However, after carrying out this procedure, $\boldsymbol{m}(\boldsymbol{r})$ no longer has the form of a propagation vector expansion, because $1/|\boldsymbol{m}_\mathrm{ic}|$ is a function of the propagation vectors $\boldsymbol{q}_i$. However, $1/|\boldsymbol{m}_\mathrm{ic}|$ has the same periodicity as the incommensurate magnetic structure and can therefore be expressed as a two-dimensional (2D) Fourier series, i.e.
\begin{equation}
	\frac{1}{|\boldsymbol{m}_\mathrm{ic}|}(k_1,k_2)=\sum_{s,t}c_{st}\exp[i(sk_1+tk_2)],
\end{equation}
where $s$ and $t$ can take all positive and negative integer values. We obtain the coefficients $c_{st}$ via a 2D discrete Fourier transform of $1/|\boldsymbol{m}_\mathrm{ic}|(\boldsymbol{r})$
We can then express $\boldsymbol{m}(\boldsymbol{r})$ as the propagation vector expansion
\begin{widetext}
\begin{equation}
	\boldsymbol{m}(\boldsymbol{r})=\Re \left[\sum_{n,p,s,t} c_{st} \boldsymbol{S}_{np} \exp[-2\pi i(\left[ n+s \right] \boldsymbol{q}_1+\left[ p+t\right] \boldsymbol{q}_2)\cdot\boldsymbol{r}]\right],
\end{equation}
\end{widetext}
which represents a weighted sum of all helices whose propagation vectors are linear combinations of $\boldsymbol{q}_1$ and $\boldsymbol{q}_2$ with integer coefficients. This series will generally be infinite, but, to keep the computational cost manageable, we restrict $-20 \le s,t \le 20$, and then discard all components whose magnitude is less than $1 \times 10^{-4}$ times the total magnetic moment.


\subsection{Contact hyperfine field}
In addition to the dipolar magnetic field, we also need to consider the contact hyperfine field at the muon site due to each of these magnetic structures. We follow the approach implemented in MuESR~\cite{doi:10.7566/JPSCP.21.011052}, whereby the hyperfine field is assumed to be isotropic and results from a scalar coupling to the magnetic moments. Each of the nearest-neighbor moments contributes to the total hyperfine field by an amount inversely proportional to the cube of its distance from the muon. This field is then scaled by the hyperfine coupling constant $A$, leading to
\begin{equation}\label{eq:hyperfine}
    \boldsymbol{B}_\mathrm{cont} = A \frac{2\mu_0}{3} \sum_{i=1}^N \frac{r^{-3}_i}{\sum_{j=1}^Nr^{-3}_j}\boldsymbol{m}_i,
\end{equation}
where $i$ (or $j$) denotes each of the $N$ nearest-neighbor Gd atoms at the muon site ($N=2$ in our case).
\subsection{Simulating $\boldsymbol{\mu^{+}\mathrm{SR}}$ spectra}
\subsubsection{Zero field}
Once we have calculated the distribution(s) of magnetic fields at the muon site(s), we can use this to predict the time-dependence of the muon-spin polarization. Solving the Larmor equation, the spin precession of the $z$ component of the muon spin subject to a magnetic field $\boldsymbol{B}$ is given by 
\begin{equation}\label{larmor}
	\frac{S_z(t)}{S}=\left(\frac{B_z}{B}\right)^2+\left[ 1-\left(\frac{B_z}{B}\right)^2 \right]\cos(\gamma_\mu B t).
\end{equation}
For a distribution of magnetic fields the total polarisation is given by
\begin{equation}\label{polarisation_distribution}
	P_z(t)=\int\frac{S_z(t)}{S} p(\boldsymbol{B})\mathrm{d}^3\boldsymbol{B}.
\end{equation}
Since our simulations give us the vector total magnetic field $\boldsymbol{B}$, we can evaluate Eq.~\eqref{polarisation_distribution} numerically as
\begin{multline}\label{polarisation_sum}
	P_z(t)=\sum_i\left(\frac{B_{z,i}}{B_i}\right)^2p_i \\+\sum_i\left[ 1-\left(\frac{B_{z,i}}{B_i}\right)^2 \right]p_i \cos(\gamma_\mu B_i t),
\end{multline}
where $p_i$ denotes the occupation probability of site $i$ and $B_{z,i}=\boldsymbol{B}_i \cdot \hat{\boldsymbol{z}}$ is the component of the local magnetic field along $z$. The index $i$ labels all of the crystallographically equivalent muon sites, each occupying different unit cells within the sample. Due to the incommensurate magnetic structure, each of these will experience a different local arrangement of magnetic moments. Note that the first term in Eq.~\ref{polarisation_sum} evaluates to a constant and does not give rise to any time dependence.
In polycrystalline samples, we have $\langle B^2_z\rangle = \langle B^2 \rangle / 3$ and hence this term gives rise to a so-called ``1/3-tail" in the polarization. However, for single-crystal samples this is not guaranteed. For our constant-moment solution, averaging over all sites gives a constant component in the polarization with value 0.294, while the higher temperature sine-II phase gives 0.293. These are broadly consistent with the ratio $A_2/(A_1+A_2)=0.306$ obtained in experiment.

 
 The fast relaxation observed in experiment is in part due to the distribution of fields that result from the incommensurate magnetic structure, but could also be partly due to dynamics. We incorporated the effect of dynamics on the polarization within the strong collision approximation, using the numerical method described in Ref.~\cite{Allodi_2014}. Taking a static polarization function $P_n = P(t_n$), defined at times $t_n = n\tau$, $0 \le n < N$, this yields a dynamic polarization function $D(\nu,t_n)=D_n$, where
 \begin{equation}
     D_n \approx \\ e^{\alpha n} \mathcal{F}^{-1} \left\{ \frac{e^{-\nu \tau} \mathcal{F} \left\{e^{-(\nu \tau+\alpha)m} P_m \right\}}{1-(1-e^{-\nu\tau})\mathcal{F} \left\{e^{-(\nu \tau + \alpha)m} P_m \right\}} \right\}_n,
 \end{equation}
 with $P_m$, $0 \le m < 2N$ being the zero-padded static relaxation function defined such that $P_m=0$ for $m \ge N$, and $\mathcal{F}$ denoting the discrete Fourier transform. The accuracy of this approximation depends on the proper tuning of the exponential weighting $\alpha$. Ref.~\cite{Allodi_2014} found that $\alpha=10/N$, with $N$ being the number of sampling points for the polarization function, led to the best accuracy when considering a Kubo-Toyabe function, and we therefore used this value for $\alpha$ in our calculations.


Taking all of these contributions into account, we model the asymmetry using a function
\begin{equation}\label{model_fit}
	A(t)=\sum_{i=1,2} A_iD_i(t;\mu_\mathrm{Gd},\nu),
\end{equation}
where the sum is over the two magnetically distinct muon stopping sites 1 that arise due to the inequivalence of $\boldsymbol{q}_1$ and $\boldsymbol{q}_2$. The internal magnetic fields are directly proportional to the magnetic moment of $\mu_\mathrm{Gd}$, which is left as a fitting parameter. Another free parameter in Eq.~\ref{model_fit} is fluctuation rate $\nu$, which is assumed to be identical at both sites. Fitting the zero-field $\mu^+$SR spectra to Eq.~\ref{model_fit} allows $\mu_\mathrm{Gd}$ and $\nu$ to be followed as a function of temperature. Fits at representative temperatures, as well as the temperature-dependence of these parameters are shown in Fig.~6 in the main text.

\begin{figure}[htb]
 	\centering
 	\includegraphics[width=\columnwidth]{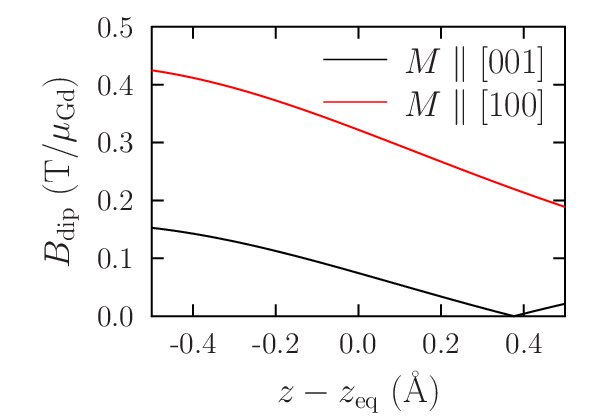}
 	\caption{Magnitude of the dipolar field at the muon site $B_\mathrm{dip}$ for a Gd moment of 1$\mu_\mathrm{B}$ at a muon site a distance $z-z_\mathrm{eq}$ along the $c$ axis from the site calculated using DFT.}
 	\label{fig:move_muon}
 \end{figure}

We note that the low-temperature moment of $4.8\mu_\mathrm{B}$ obtained from our model is $\approx$30\% smaller than the accepted value of 7.0$\mu_\mathrm{B}$. Small inaccuracies in the muon stopping site could result in changes to the dipolar field at the muon site for a given moment size, which would also change our estimate for the hyperfine coupling $A$ obtained from our modelling. 
Following the effects of these inaccuracies through to the moment obtained by our modelling in not straightforward, particularly in light of the complicated incommensurate magnetic structure. To get an idea for the magnitude of the errors that could be introduced, we studied the effect of changing the muon position on the dipolar part of the magnetic field at the muon site for the field-polarized ferromagnetic structures considered in our analysis of the TF $\mu^+$SR spectra. Taking into account the symmetry of the muon site, we considered displacements of a muon at (0.50, 0.00, 0.07) along the $c$ axis. The resulting variation of the dipolar fields is shown in Fig.~\ref{fig:move_muon}. Here a decrease in the dipolar field at the muon site corresponds to an increase in the estimate of the Gd moment. A 30\% decrease in $B_\mathrm{dip}$ from its value at $z_\mathrm{eq}$ requires a displacement of 0.11 \AA{} for magnetization $\boldsymbol{M} \parallel [001]$ or 0.35 \AA{} for $\boldsymbol{M} \parallel [100]$. Both of these displacements are significantly larger than the tolerance for the atomic positions required in our structural relaxations, with the latter being well outside any errors in atomic distances that one might expect to observe using DFT. Inaccuracies in the muon position from DFT are therefore unlikely to be solely responsible for the discrepancy between our extracted Gd moment and the accepted value.  Note that manually moving the muon position in this manner meant that we were unable to take into account the muon-induced distortions of the nearby Gd atoms. 

Another possible contributor to this error is the need to estimate a value for the demagnetizing factor $N$. While the value of $N$ does not affect the dipolar field due to the constant-moment solution directly, since this magnetic structure has no net magnetization, it does affect our estimate for the hyperfine coupling constant $A$. To get a handle on the possible size of such error, we consider the fact that in our modelling of the field-polarized state, $A$ and $N$ always appear in the combination $-0.140 N + 7.77A$, and we can therefore write our estimate for $A$ as a function of $N$ as $A(N)=A(N=1)+0.018N$. Given the fact $A(N=1)=-0.0040(9)$ \AA$^{-3}$, our estimate of $A$ therefore depends quite strongly on $N$; it becomes more negative as $N$ decreases from one. Intuitively, one might expect a larger magnitude of $A$ (as would result from $N < 1$) might lower our estimate for $\mu_\mathrm{Gd}$, suggesting that our choice of $N=1$ cannot be the reason for our underestimate. However, the vectoral addition of magnetic field contributions at the muon site, in conjunction with the complicated nature of the magnetic structure, prevents us from making a definitive conclusion in this regard.

A further possible source of error is the approximations made in incorporating the hyperfine field. The approach we have used is rather crude, and the quantum nature of the hyperfine interaction requires knowledge of the electronic spin density in the vicinity of the muon. While our DFT calculations provide this in principle, we note that the hyperfine coupling constant obtained using DFT show poor agreement with those obtained from our fitting. This is likely due to the failure of our calculation to properly account for the physics of the RKKY interaction. However, doing so is beyond the scope of the present study.

\subsubsection{Transverse field}
Simulating the TF $\mu^+$SR spectra starts in much the same way as for the zero-field spectra, though we need to account for the fact that the magnetic field $\boldsymbol{B}$ also bears a contribution from the applied field, i.e., $\boldsymbol{B} = \boldsymbol{B}_\mathrm{int}+ \boldsymbol{B}_0$. Conventionally, we measure $P_x(t)$ in a transverse-field experiment. This has the same form as Eqs.~\eqref{larmor}-\eqref{polarisation_sum} after substituting $z \to x$. With this convention, the applied field is then parallel to the $z$ axis. For the measurements with $\boldsymbol{B}_0 \parallel [001]$, the crystallographic $c$ axis of the crystal is parallel to $z$, and the $a$ axis coincides with $\hat{\boldsymbol{x}}$. Simulating the spectra for $\boldsymbol{B}_0 \parallel [100]$ requires us to rotate the internal fields (equivalent to rotating the sample), such that the $a$ axis is parallel to $z$, with the $c$ axis parallel to $\hat{\boldsymbol{x}}$. The HAL spectrometer comprises eight detectors in a ring. We therefore wish to consider the muon-spin polarization along the directions $\mathbf{X}_j$, where
\begin{equation}
    \mathbf{X}_j = \hat{\boldsymbol{x}} \cos (j\theta) + \hat{\boldsymbol{y}} \sin (j\theta),
\end{equation}
with $\theta = \pi/4$ and $0 \le j < 4$. We then calculate the average field spectrum for all of these detectors, which, following the linear property of Fourier transforms, is obtained from the Fourier-transform of the average of these polarizations,
 \begin{equation}
 	P(B) = \mathcal{F} \left\{ \frac{1}{4} \sum_{j=0}^3 P_{X_j}(t) \right\}.
 \end{equation}
  
\section{Magnetic susceptibility}\label{sec:magnetometry}

\begin{figure}[htb]
 	\centering
 	\includegraphics[width=0.89\columnwidth]{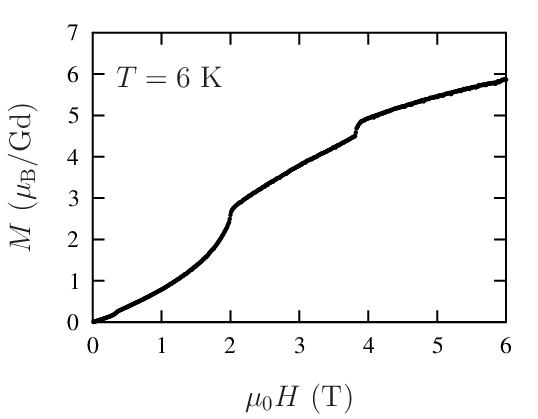}
 	\caption{Magnetization as a function of applied field \mbox{$\boldsymbol{B_0}=\mu_0\boldsymbol{H} \parallel [100]$} at $T=6$~K.}
 	\label{fig:MvH}
 \end{figure}

\begin{figure}[htb]
 	\centering
 	\includegraphics[width=0.89\columnwidth]{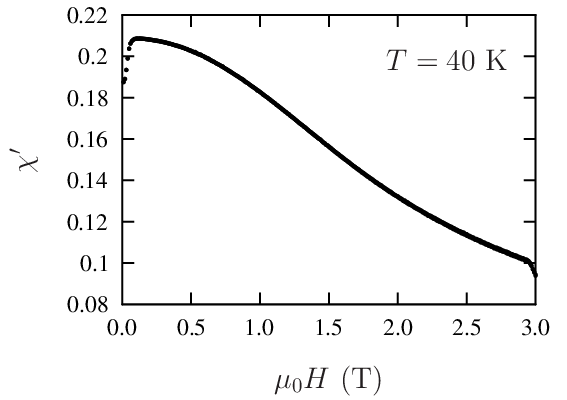}
 	\caption{Real part $\chi'$ of the ac magnetic susceptibility as a function of applied field $\boldsymbol{B}_0=\mu_0\boldsymbol{H} \parallel [100]$ at $T=40$~K.}
 	\label{fig:ac}
 \end{figure}

We measured the magnetic properties of a 127 mg single crystal, with fields applied perpendicular to its large (001) face, using a Quantum Design MPMS3 magnetometer. The crystal was secured in the brass trough sample holder, with the magnetic field applied in-plane, i.e. along [100]. Measurements of the dc moment as a function of temperature between 2 and 60 K were made in applied fields of $\mu_0H = 0, 0.1, 0.25$ and 0.5~T. The value of the moment at 55 K from the scan in $\mu_0 H=0.5$~T was used for comparison with the moment obtained from our analysis of the angle-dependent muon Knight shift measurements. These were followed by a series of field scans carried out in 2-K intervals between 2 and 50 K. Both the dc moment (using the VSM technique) and ac susceptibility were measured at each field and temperature. The magnetization as a function of applied field at $T=6$~K, $M(H)$, is shown in Fig.~\ref{fig:MvH}, and is used to fix the ferromagnetic component of the magnetic structures used in our dipolar field calculations. For the ac measurements we used an oscillating field with amplitude 10 Oe and frequency 113 Hz. From the real part $\chi'$ of the ac susceptibility, we constructed the $B-T$ phase diagram shown in Fig. 3(j) in the main text. We show here in Fig.~\ref{fig:ac} $\chi'$ as a function of applied magnetic field at 40 K. This data forms part of Fig. 3(j) in the main text, but is reproduced here to highlight the initial increase in $\chi'$ with field, that plateaus by around 0.1 T. This is followed by a monotonic decrease until around 3 T, highlight the absence of a phase transition between 0.5 and 1 T for this orientation.

\newpage
\bibliography{references}